\documentclass[%
 aip,
 amsmath,amssymb,
 reprint,%
]{revtex4-1}

\usepackage{indentfirst}
\usepackage{tablefootnote}

\usepackage{graphicx}
\usepackage{dcolumn}
\usepackage{bm}

\usepackage[utf8]{inputenc}
\usepackage[T1]{fontenc}
\usepackage{mathptmx}
\usepackage{etoolbox}

\usepackage[dvipsnames]{xcolor}
\usepackage{caption}

\usepackage[colorlinks=true,
            linkcolor=blue,
            urlcolor=blue,
            citecolor=blue]{hyperref}
\usepackage{adjustbox}

\usepackage[version=3]{mhchem} 

\newcommand*{\blauw}[1]{\textcolor{blue}{#1}}

\preprint{AIP/123-QED}
\makeatletter
\def\@email#1#2{%
 \endgroup
 \patchcmd{\titleblock@produce}
  {\frontmatter@RRAPformat}
  {\frontmatter@RRAPformat{\produce@RRAP{*#1\href{mailto:#2}{#2}}}\frontmatter@RRAPformat}
  {}{}
}%
\makeatother

\begin{document}

\title{A Modular and Extensible CHARMM-Compatible Model for All-Atom Simulation of Polypeptoids}

\author{Alex Berlaga}\thanks{These authors contributed equally to this work.}
\affiliation{ 
Department of Chemistry, University of Chicago, Chicago, Illinois 60637, United States of America
}

\author{Kaylyn Torkelson}\thanks{These authors contributed equally to this work.}%
\affiliation{ 
Department of Chemical Engineering, University of Washington, Seattle, Washington 98195, United States of America
}

\author{Aniruddha Seal}
\affiliation{%
Department of Chemistry, University of Chicago, Chicago, Illinois 60637, United States of America
}
\author{Jim Pfaendtner}
\email{pfaendtner@ncsu.edu}
\affiliation{Department of Chemical and Biomolecular Engineering, North Carolina State University, Raleigh, North Carolina 27695, United States of America}

\author{Andrew L.\ Ferguson}
\email{andrewferguson@uchicago.edu}
\affiliation{Pritzker School of Molecular Engineering, University of Chicago, Chicago, Illinois 60637, United States of America}
\affiliation{ 
Department of Chemistry, University of Chicago, Chicago, Illinois 60637, United States of America
}

\date{\today}

\clearpage
\newpage

\begin{abstract}

\noindent Peptoids (N-substituted glycines) are a class of sequence-defined synthetic peptidomimetic polymers with applications including drug delivery, catalysis, and biomimicry. Classical molecular simulations have been used to predict and understand the conformational dynamics of single peptoid chains and their self-assembly into diverse morphologies including sheets, tubes, spheres, and fibrils. The CGenFF-NTOID model based on the CHARMM General ForceField has demonstrated success in enabling accurate all-atom molecular modeling of the structure and thermodynamic behavior of peptoids. Extension of this force field to new peptoid side chain chemistries has historically required parameterization of new side chain bonded interactions against \textit{ab initio} and/or experimental data. This fitting protocol improves the accuracy of the force field but is also burdensome and time consuming, and precludes modular extensibility of the model to arbitrary peptoid sequences. In this work, we develop and demonstrate a Modular Side Chain CGenFF-NTOID (MoSiC-CGenFF-NTOID) as an extension of CGenFF-NTOID employing a modular decomposition of the peptoid backbone and side chain parameterizations wherein arbitrary side chain chemistries within the large family of substituted methyl groups (i.e., \ce{-CH_3}, \ce{-CH_2R}, \ce{-CHRR$^{\prime}$} \ce{-CRR$^{\prime}$R$^{\prime\prime}$}) are directly ported from CGenFF without any additional reparameterization. We validate this approach against \textit{ab initio} calculations and experimental data to to develop a MoSiC-CGenFF-NTOID model for all 20 natural amino acid side chains along with 13 commonly-used synthetic side chains, and present an extensible paradigm to efficiently determine whether a novel side chain can be directly incorporated into the model or whether refitting of the CGenFF parameters is warranted. We make the model freely available to the community along with a tool to perform automated initial structure generation. We anticipate that these tools will help enable high-throughput virtual screening and simulation campaigns and advance engineering and design efforts of peptoid biomaterials.
\end{abstract}
\maketitle

\clearpage
\newpage

\section{Introduction}

\begin{figure}[ht]
    \centering
    \includegraphics[trim={0 0 0 0}, clip, scale=0.35]{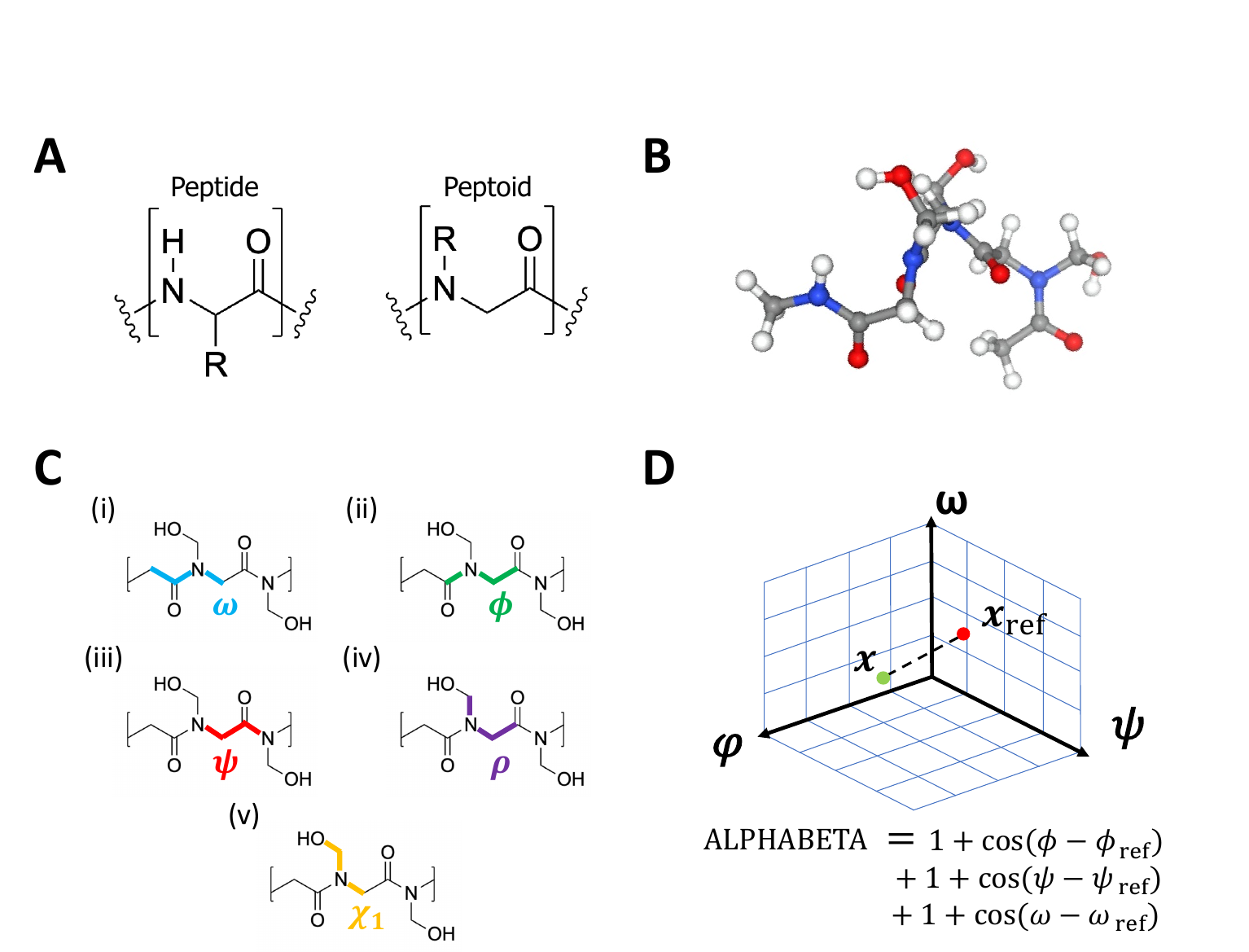}
    \caption{Illustration of peptoid structure and chemistry. (A) Comparison of the chemical structure of peptides and peptoids. (B) Ball-and-stick molecular model of a serine analogue peptoid trimer (NSer--NSer--NSer). Visualization constructed using NGLView\cite{10.1093/bioinformatics/btx789}. (C) Illustration of the (i) $\omega$, (ii) $\phi$, (iii)  $\psi$, (iv) $\rho$, and (v) $\chi_1$ dihedral angles in the context of a NSer peptoid dimer. The elimination of the partial double bond character of the planar amide bond means that rotations of the $\omega$ dihedral are more accessible in peptoids compared to peptides and, depending on the side chain chemistry, permits access to both the cis and trans configurations. The $\rho$ dihedral is closely related to the $\phi$ dihedral and is absent in peptides. (D) Schematic illustration of the alphabeta collective variable\cite{Tribello_2014, BONOMI20091961} used to measure peptoid conformational similarity as a distance in the three-dimensional space of the $\phi$, $\psi$, and $\omega$ dihedral angles. 
    }
    \label{fig:peptoid-basics}
\end{figure}

Peptoids, or N-substituted glycines, are a class of synthetic, biomimetic sequence-defined polymers. Peptoids are chemically similar to peptides but differ in the side chains being bonded to the backbone nitrogen instead of the backbone $\alpha$-carbon, which endows them with distinct behaviors and properties \cite{Sun2013, C3BM60269A, Gangloff2016} (Figure \ref{fig:peptoid-basics}A,B). In particular, the elimination of the chiral center at the $\alpha$-carbon, elimination of the ---NH hydrogen bond donor, and stabilization of the cis configuration of the backbone $\omega$ dihedral endow peptoids with significantly more conformational flexibility than peptides\cite{Sun2013, zuckermann2011, yan2018controlled, Xuan2019, rosales2010control} and the emergence of distinct folding patterns\cite{Lee2005, Gangloff2016, Kirshenbaum1998, Butterfoss2009} and engineerable secondary structural elements\cite{Sun2013, SHI2023101677, C8SC04240C, C3BM60269A, doi:10.1073/pnas.0708254105, Eastwood2023} than those that exist for proteins and peptides. Peptoids are also easily synthesizable by iterative solid-phase synthesis \cite{C3BM60269A, Sun2013} and resistant to proteolysis \cite{Schunk2023}. Taken together, these attributes have led to numerous applications for peptoids in biological and materials engineering, such as chelators of metals and multimetallic clusters\cite{Sun2013, C8SC04240C, C9SC01068H,molecules23020296}, protein and antibody mimics \cite{C3BM60269A}, anti-microbial agents\cite{Lin2016, doi:10.1073/pnas.0708254105, Radhe2021, Lin2022, Nielsen2022}, and selective toxin detectors\cite{Wang2020}. 

Computational prediction of sequence-dependent peptoid structure and dynamics is valuable for understanding, optimization, engineering, and design of the properties and functions of peptoid molecules and materials in applications ranging from drug delivery to enzyme design\cite{Jian2022, C8SC04240C, Rosalba2023}. Classical all-atom molecular dynamics (MD) simulations have been used to model peptoid structure and dynamics and provide molecular-level rationalizations for experimental observations of stability, reactivity, and other physicochemical properties. For many years, general-purpose biomolecular force fields such as CGenFF\cite{Vanommeslaeghe2010-ao}, OPLS\cite{Jorgensen1996}, GAFF\cite{Wang2004}, and GROMOS\cite{Scott1999}, have been available for the all-atom simulation of arbitrary polynucleotide or polypeptide sequences composed of naturally occurring subunits. These force fields have contributed to important findings in the fields of enzyme kinetics\cite{Aqvist2020}, protein folding\cite{Daidone2003-vb}, peptide drug discovery\cite{EVERS20191404}, DNA flexibility\cite{HEDDI2010123}, as well as countless other applications. Conversely, there exists no general-purpose force field for peptoid simulation. Previous studies that have applied peptide-centric force fields such as GAFF\cite{Wang2004} and CGenFF to peptoid systems and have revealed that those parameters may often lead to striking inaccuracies in conformational free energy landscapes \cite{https://doi.org/10.1002/jcc.25850, Voelz2010, https://doi.org/10.1002/jcc.23478, mukherjee2015}. As a result, it is currently common practice to parameterize bespoke force field parameters for each peptoid residue of interest by, typically, bottom-up fitting against \textit{ab initio} calculations \cite{https://doi.org/10.1002/jcc.25850, Jain2023, Harris2023}. To date, peptoid-specific force field parameters have been developed for the GAFF\cite{Harris2023,mukherjee2015, Tuttle2023}, AMBER99SB \cite{Prakash2018}, CHARMM\cite{https://doi.org/10.1002/jcc.23478, https://doi.org/10.1002/jcc.25850, Zhao2020, Mannige2015, Alamdari2023, Jain2023}, OPLS\cite{Du2018}, and DREIDING\cite{Hoyas2018} force fields, but these parameterizations requires significant computational resources and manual tuning. A generic peptoid force field that provides accurate predictions for commonly used peptoid side chains of interest would be an enabling tool for high-throughput computational screening for the discovery, engineering, and optimization of peptoids with desired properties.

Weiser and Santiso recently developed a peptoid-centric version of the CHARMM General ForceField (CGenFF) by conducting \textit{ab initio} calculations of the potential energy landscape of a capped peptoid monomer with a methyl (NAla) side chain and refitting the $\rho$, $\psi$, and $\omega$ backbone dihedrals (Figure \ref{fig:peptoid-basics}C) to develop a transferable parameterization scheme suitable for the large family of peptoid side chains attached to the backbone via methylene or substituted methylene bridges \cite{https://doi.org/10.1002/jcc.25850}. All other backbone bonded and non-bonded parameters were adopted directly from CGenFF, with the exception of the peptoid nitrogen for which a new NTOID atom type was defined. Henceforth, we will refer to this force field as CGenFF-NTOID. Two additional side chains -- s-(1)-phenylethyl (Nspe) and phenylmethyl (NPhe)  -- were parameterized and added to the model by explicitly refitting the $\chi_1$ dihedral, which controls the rotation of the side chain relative to the peptoid backbone (Figure \ref{fig:peptoid-basics}C), against \textit{ab initio} calculations.
In this work, we hypothesize that the family of side chains attached to the backbone via methylene or substituted methylene bridges can be accurately modeled using CGenFF-NTOID for the backbone (i.e., CGenFF with refitted $\rho$, $\psi$, and $\omega$ backbone dihedrals) and default CGenFF parameters for the side chain. This encompasses all side chains belonging to the family of substituted methyl groups (i.e., \ce{-CH_3}, \ce{-CH_2R}, \ce{-CHRR$^{\prime}$} \ce{-CRR$^{\prime}$R$^{\prime\prime}$}), and where we have implicitly asserted that the methyl carbon remains sp\textsuperscript{3} hybridized and that none of the substituents introduce a bonded connection to the backbone. By adopting this hypothesis of modularity between the side chain and backbone parameterizations, we can eschew refitting of any side chain parameters (including the $\chi_1$ dihedral) away from the default CGenFF parameterization, and establish an extensible force field capable to modeling arbitrary peptoid sequences comprising substituted methyl group side chains. 
We refer to this model and paradigm as Modular Side Chain CGenFF-NTOID (MoSiC-CGenFF-NTOID).

We test our hypothesis by considering a selection of 34 commonly used and chemically diverse peptoid side chains. We test the accuracy of the MoSiC-CGenFF-NTOID predictions in a hierarchy of validations starting at the level of individual residues and escalating to the behavior of self-assembled supramolecular aggregates. First, we compare Ramachandran potential energy distributions against those produced by \textit{ab initio} calculations. Second, we compare cis/trans isomerization energies and free energies against \textit{ab initio} calculations and experimental measurements. Third, we compare the predictions of our model for the self-assembly of minimal peptoid sequences to experimental observations of their aggregation states \cite{Castelletto2020}. We find that 25/34 side chains exhibited performance similar or superior to that of the three side chains -- methyl (NAla), s-(1)-phenylethyl (Nspe) and phenylmethyl (NPhe)  -- explicitly parameterized against \textit{ab initio} data and reported in the original CGenFF-NTOID paper\cite{https://doi.org/10.1002/jcc.25850} (Figure \ref{fig:all-residues}). We find 8/34 additional side chains to show satisfactory performance, but could benefit from additional parameterization. Only 1/34 side chains show sufficiently poor performance that we do not recommend use prior to a reparameterization. The single side chain showing poor performance (Nph) was constructed as a control case and is the only side chain considered that does not either belong to the family of substituted methyl groups or for which a bespoke reparameterization was available. These results provide strong support for our hypothesis that the parameters for a variety of peptoid side chains within the family of substituted methyl groups can be adopted directly from CGenFF-NTOID without additional reparameterization. We also present a protocol to assess whether the accuracy is satisfactory and, if not, motivate refitting of the $\chi_1$ dihedral using the procedure developed by Weiser and Santiso\cite{https://doi.org/10.1002/jcc.25850}.

Given these encouraging results, we developed a publicly-available open-source Python package implementing MoSiC-CGenFF-NTOID built on the CGenFF-NTOID model\cite{ https://doi.org/10.1002/jcc.23478, https://doi.org/10.1002/jcc.25850}. We developed our model for CGenFF based on the Feburary 2021 version of CHARMM36\cite{Huang2013-wx}, but the model can be extensibly and straightforwardly updated for compatibility with future CHARMM releases. Our model implements the methyl (NAla), s-(1)-phenylethyl (Nspe), and phenylmethyl (NPhe) reparameterizations explicitly refitted against \textit{ab initio} data by Weiser and Santiso, as well as the same reparameterizations for any side chains with a $\chi_1$ dihedral containing identical atom types to one of the aforementioned side chains. All other side chains employ the default CGenFF parameterizations originally developed for peptides and proteins\cite{Vanom2012-ku, Vanommeslaeghe2012, Vanommeslaeghe2010-ao}. We note that the Gly and Pro peptoid residues are chemically identical to the corresponding peptide residues and we directly employ the peptide parameterization within CGenFF.
The package also provides Python scripts to generate molecular structures of arbitrary length peptoids based on user-defined dihedral conformations, including the twelve common rotameric states\cite{Voelz2010}, suitable for use as initial states from which to launch MD simulations. Taken together, our package furnishes a generic and extensible all-atom force field for the family of peptoid side chains comprising substituted methyl groups. The model currently contains parameterizations for those side chains considered in this work, but it is straightforward for users to add their own side chains not contained in this set. Although our results provide strong support for the addition of any side chain in the family of substituted methyl groups without any additional reparameterization, we advise validating the parameterization by, at a minimum, comparing Ramachandran potential energy plots against those generated by \textit{ab initio} calculations to determine whether additional reparameterization may be warranted. We anticipate that the MoSiC-CGenFF-NTOID may prove valuable to the community in enabling high-throughput computational screening and engineering of peptoid structure, dynamics, and properties.

\section{Methods}

\subsection{MoSiC-CGenFF-NTOID Parameterization and Residue Topology Creation}

The MoSiC-CGenFF-NTOID is a straightforward extension of the all-atom CGenFF-NTOID model for peptoids developed by Weiser and Santiso \cite{https://doi.org/10.1002/jcc.25850}. The backbone parameters are taken directly from CGenFF-NTOID \cite{https://doi.org/10.1002/jcc.25850} and the side chain parameters are adopted directly from the CGenFF model developed for peptides and proteins \cite{Vanommeslaeghe2010-ao}. The underlying hypothesis of MoSiC-CGenFF-NTOID, and really the sole innovation over CGenFF-NTOID, is the hypothesis of a backbone--side chain modular decomposition, which positions MoSiC-CGenFF-NTOID as a generic and extensible peptoid model capable of treating any side chain in the family of substituted methyl groups (i.e., \ce{-CH_3}, \ce{-CH_2R}, \ce{-CHRR$^{\prime}$} \ce{-CRR$^{\prime}$R$^{\prime\prime}$}), and where we have implicitly asserted that the $\beta$-carbon remains sp\textsuperscript{3} hybridized and that none of the substituents introduce a bonded connection to the backbone. 

\begin{figure}
    \centering
    \includegraphics[scale=0.22]{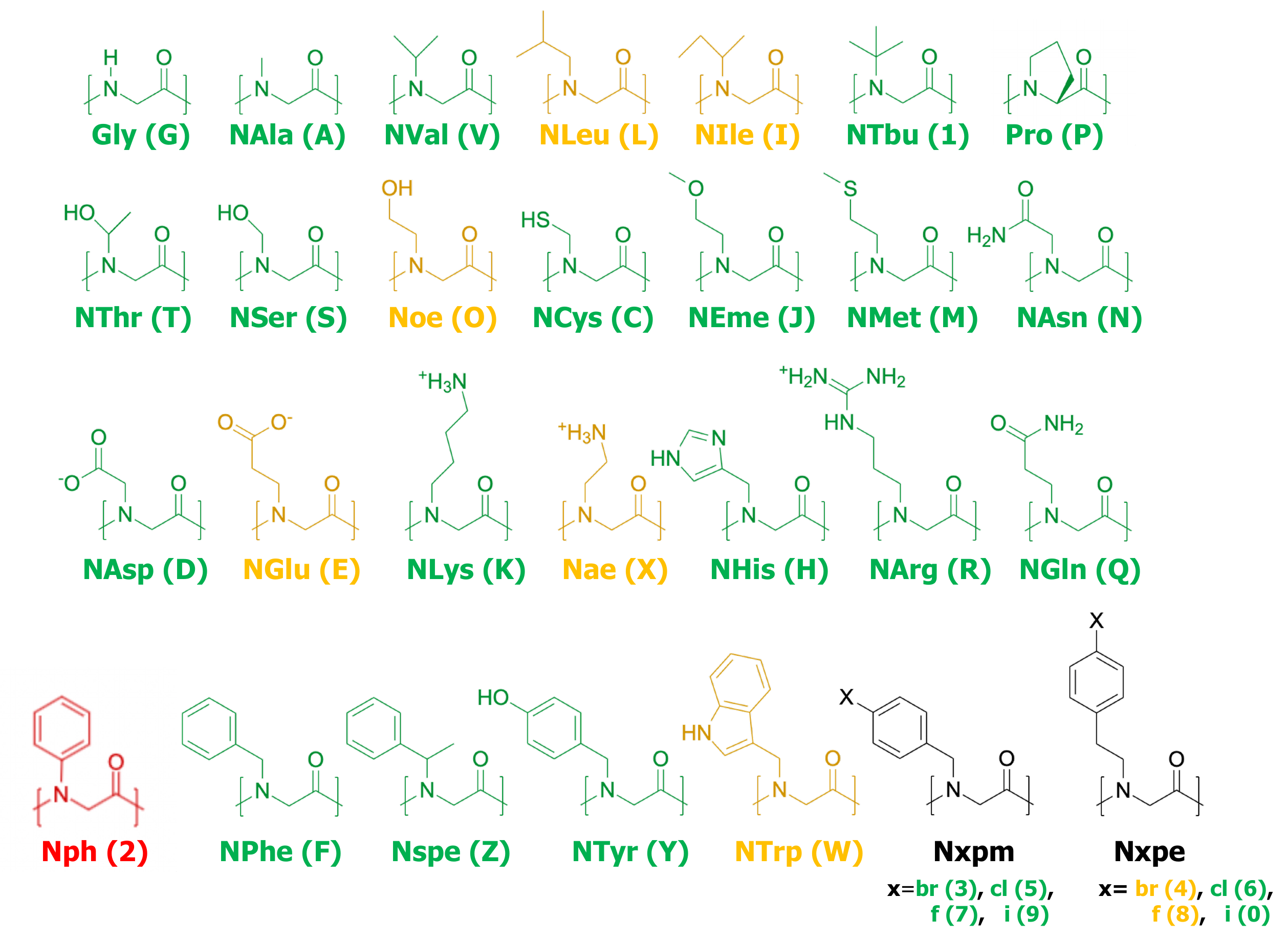}
    \caption{Illustration of the 34 side chains included in our initial release of MoSiC-CGenFF-NTOID, including all 20 naturally occurring amino acid side chains and 14 other commonly studied side chains. Side chains determined to produce good agreement with \textit{ab initio} Ramachandran plot projections and cis/trans potential energy differences are colored green, those with fair performance orange, and those with poor performance red. Each side chain is labeled with its abbreviation and single letter or single number code (cf.\ \blauw{Table S1}). 
    }
    \label{fig:all-residues}
\end{figure}

We explicitly test this hypothesis by developing an initial MoSiC-CGenFF-NTOID implementation comprising the 34 peptoid side chains illustrated in Figure \ref{fig:all-residues} and detailed in \blauw{Table S1}. This set comprises all 20 naturally occurring amino acid side chains, as well as 14 other side chains commonly studied in the experimental and computational literature\cite{Monahan2022, Li2021, Eastwood2023}. All side chains, with the exception of Gly, Pro, and Nph belong to the family of substituted methyl groups. As mentioned above, Gly and Pro are chemically identical to their peptide analogs and so their parameterizations are contained within CGenFF, and Nph was considered as a negative control. We treat all backbone bonded interactions based on the CGenFF-NTOID model developed by Weiser and Santiso corresponding to a CGenFF force field with refitted $\rho$, $\psi$, and $\omega$ backbone dihedrals\cite{https://doi.org/10.1002/jcc.25850}. Partial charges, non-bonded parameters, and side chain bonded interactions are all adopted directly from CGenFF \cite{Vanom2012-ku, Vanommeslaeghe2012, Vanommeslaeghe2010-ao}, with the exception of NAla, Nspe, and NPhe, and sidechains containing identical $\chi_1$ atom types, which were previously reparameterized against \textit{ab initio} calculations by Weiser and Santiso \cite{https://doi.org/10.1002/jcc.25850}. 

\subsection{Structure Generation} \label{subsec:strucgen}


We complement the MoSiC-CGenFF-NTOID model with a peptoid structure generator that is capable of producing a PDB structure of arbitrary peptoid sequences with a defined set of  $\phi$, $\psi$, and $\omega$ dihedral angles. The structure generator constructs an initial polyglycine peptide backbone in the desired dihedral angle conformation using PeptideBuilder\cite{tien2013peptidebuilder}, then transforms it into the desired peptoid sequence by grafting the desired sequence of side chains to the backbone \ce{N} atoms, adding an acetyl (\ce{-C(O)CH3}) N-terminal cap, and one of three possible C-terminal caps: amino (\ce{-NH2}), aminomethyl (\ce{NHMe}), or aminodimethyl (\ce{-N(Me)2}). The atom types of the resulting PDB file are modified for compatibility with the MoSiC-CGenFF-NTOID model, and a backbone-restrained, steepest-descent energy minimization conducted to eliminate and high-energy steric clashes\cite{ABRAHAM201519}.

\subsection{Classical Molecular Dynamics Simulations}

\subsubsection{All-atom Molecular Dynamics Simulations of Peptoids \label{subsec:AAMD}} 

All-atom MD simulations of various peptoid sequences were conduced under the MoSiC-CGenFF-NTOID model using Gromacs 2021.4\cite{ABRAHAM201519}. In single chain simulations that probed intramolecular properties and statistics, an acetyl (Ac) N-capped and aminomethyl (NHMe) C-capped peptoid monomer or polymer -- generally a dimer, trimer, or pentamer -- was, respectively, placed into a 3.5$\times$3.5$\times$3.5 nm$^3$ or 6.0$\times$6.0$\times$6.0 nm$^3$ cubic box implementing periodic boundary conditions in all dimensions. For simulations of multiple chains used to probe multi-chain aggregation and supramolecular self-assembly, we randomly placed 50 end-capped trimer peptoids into a 6.0$\times$6.0$\times$6.0 nm$^3$ cubic box. All systems were energy-minimized using the steepest descent algorithm for a maximum of 100,000 steps, or until the maximum force was below 1000 kJ/mol.nm. Systems were solvated in SPC/E water\cite{spce} to a density of 1 g/cm\textsuperscript{3}, corresponding to conditions of standard temperature and pressure, and corresponding to $\sim$850-860 water molecules for the single chain systems in the 3.5 nm cubic box, $\sim$7000-7020 for the single-chain systems in a 6.0 nm cubic box, and $\sim$5500-5900 for the multiple chain systems. In the case of charged systems, compensatory Na$^+$ or Cl$^-$ ions were added as necessary to neutralize the net charge. Electrostatics were treated using Particle-Mesh-Ewald \cite{PME} with a real-space cutoff of 1.0 nm and a Fourier grid spacing of 0.08 nm that was subsequently optimized during runtime. Lennard-Jones interactions were smoothly shifted to zero at a cutoff of 1.0 nm. Initial velocities were assigned from a Maxwell-Boltzmann distribution at 300 K. The classical equations of motion were integrated using a 2 fs time step under the leap-frog algorithm \cite{birdsall2018plasma}. Covalent bonds involving hydrogen atoms were fixed using the LINCS algorithm\cite{LINCS}. Center-of-mass (COM) translation and rotation were removed every 1 ps. A Verlet cutoff scheme was used with a neighbor list updated every 15 time steps. Periodic boundary conditions were applied, with a cut-off distance of 1.0 nm for neighbor searching. Equilibration was first performed for 200 ps in the NVT ensemble at 300 K using a Bussi-Parrinello-Donadio velocity rescaling thermostat\cite{Bussi2007} with a time constant of 0.1 ps. This was followed by 200 ps of equilibration in the NPT ensemble at 300 K and 1 bar with temperature maintained using a Bussi-Parrinello-Donadio velocity rescaling thermostat\cite{Bussi2007} with a coupling time constant of 0.1 ps and an isotropic Berendsen barostat with a coupling time constant of 1.0 ps and an isothermal compressibility of 4.5$\times$10$^{-5}$ bar$^{-1}$. Production runs were conducted in the NPT ensemble at 300 K and 1 bar employing a Nos\'e-Hoover thermostat\cite{Evans1985} with a time constant of 1.0 ps and an isotropic Parinello-Rahman barostat\cite{ParinelloRahman} with a time constant of 1.0 ps and an isothermal compressibility of 4.5$\times$10$^{-5}$ bar$^{-1}$. Simulation trajectories were saved at a period of 2.0 ps. Simulations were conducted on 1$\times$NVIDIA V100 or A100 GPU cards to achieve execution speeds of $\sim$50-700 ns/day for the single chain systems and $\sim$200-500 ns/day for the multiple chain systems. Example input files for the single chain and multiple chain runs are provided in the MoSiC-CGenFF-NTOID GitHub repository at \url{https://github.com/UWPRG/mftoid-rev-residues}.

\subsubsection{Enhanced Sampling Calculations Using Well-Tempered Parallel Bias Metadynamics}\label{subsec:bias}


It is well known that enhanced sampling is typically required to accelerate structural transitions in the peptoid backbone dihedrals -- most pressingly cis/trans isomerizations of the $\omega$ dihedral but also rotations around the $\phi$ and $\psi$ angles (Figure \ref{fig:peptoid-basics}C) -- that separate states of similar thermodynamic stabilities by high free energy barriers\cite{https://doi.org/10.1002/jcc.23478, https://doi.org/10.1002/jcc.25850}. To accelerate sampling of the thermally-accessible configurational space for both our end-capped monomers and trimers for comparison against \textit{ab initio} calculations of the Ramachandran potential energy distributions, we employ well-tempered parallel bias metadynamics (WT-PBMetaD) \cite{Pfaendtner2015, PhysRevLett.104.190601, Zhao2020} coupled to the following collective variables: (i) all $\phi$, $\psi$, and $\omega$ backbone dihedrals, (ii) the molecular radius of gyration ($R_g$), and (iii) the alphabeta distances\cite{Tribello_2014, BONOMI20091961}, $ALPHABETA = (1 + \cos (\phi - \phi_{ref})) + (1 + \cos(\psi - \psi_{ref})) + (1 + \cos(\omega - \omega_{ref}))$ measuring the configurational similarity of each contiguous $\{ \phi, \psi, \omega \}$ triplet from the twelve stable peptoid backbone minima reported in Ref.\ \cite{Alamdari2023} and listed in \blauw{Table S2} (Figure \ref{fig:peptoid-basics}D). For our end-capped monomers, this set comprises 16 coupled collective variables, and for our end-capped trimers, 22 coupled collective variables. To confine the exploration to a chemically relevant region of conformational space, minimum and maximum allowed values were defined for each CV adapted from a previous peptoid simulation study\cite{Alamdari2023}: [0, 3] nm for $R_g$, [($-\pi$), $\pi$] radians for each dihedral angle, and [(-1), 50] for the alphabeta distances.

Within the WT-PBMetaD calculations, we employed a bias factor of $\gamma$ = 20-30, an initial Gaussian height of $W_0$ = 3-5 kJ/mol, and a Gaussian width of $\sigma$ = 0.1 nm for $R_g$, 1 for the alphabetas, and 0.35 radians for the dihedrals. We employed a deposition rate of 1 ps\textsuperscript{-1}. Each system is simulated using the GROMACS simulation engine\cite{ABRAHAM201519} patched with the PLUMED 2.7.2\cite{BONOMI20091961, Tribello_2014} enhanced sampling libraries. Calculations of end-capped monomers and trimers conducted on 1$\times$NVIDIA V100 or A100 GPU cards achieved execution speeds of 50-300 ns/day. Convergence of the WT-PBMetaD calculations was assessed following best practices \cite{bussi2020using} by monitoring decay in the Gaussian hill heights deposited along the coupled collective variables, tracking the free diffusion in these collective variables, and checking for frequent cis/trans transitions in the $\omega$ dihedrals. We define convergence to have been reached when the Gaussian hill heights drop to less than 5\% of the initial hill height, we observe free diffusion in all coupled CVs, and we achieve more than 20 cis/trans transitions in the $\omega$ dihedrals. These criteria are met for all peptoids considered in this work within 500 ns of simulation time. An illustrative example of the monitored convergence criteria for the Nclpm trimer is presented in \blauw{Figure S1}. Once converged, the biased trajectories were reweighted to the unbiased ensemble using the Torrie-Valleau reweighting method \cite{TORRIE1977187}.

\subsubsection{Classical and \textit{Ab Initio} Potential Energy Calculations}\label{subsec:energy}



Potential energies in vacuum were computed for identical peptoid structures using both quantum mechanical and classical mechanical methods to compare the predictions of the MoSiC-CGenFF-NTOID model to \textit{ab initio} calculations within Ramachandran projections and calculations of the cis/trans potential energy difference $\Delta U_{\text{cis/trans}}$. We generated acetyl N-capped and aminomethyl C-capped monomers of each of the 34 peptoid monomers incorporated in our initial force field (Figure \ref{fig:all-residues}) and simulated for 500 ns under vacuum conditions at 300 K to generate a library of configurations for which to calculate potential energies. To eliminate artifacts due to non-charge neutral systems and compatibility with \textit{ab initio} calculations \cite{KONERMANN2018104}, simulations were conducted in a pseudo-infinite cubic box with side length 999.9 nm and real-space cutoffs for the Lennard-Jones and Coulomb interactions of 333.3 nm. Electrostatics were treated using a plain cutoff of 333.3 nm. Temperature was maintained using a Nose-Hoover thermostat. Otherwise, all other simulation parameters and biased collective variables were identical to those reported in Sections \ref{subsec:AAMD} and \ref{subsec:bias}. Large temperature fluctuations are to be expected for such a small system and the temperature control is not expected to be be terribly tight, but this is not material for the present test which need only generate a diverse sampling of molecular configurations for the purposes of classical and quantum potential energy comparisons. Structures were harvested every 2 ps.

Classical mechanical potential energies under the MoSiC-CGenFF-NTOID model were computed using the Gromacs $\texttt{energy}$ command. Snapshots were then classified as cis ($-\frac{\pi}{2} < \omega \leq \frac{\pi}{2}$) or trans ($[-\pi < \omega \leq -\frac{\pi}{2}] \cup [\frac{\pi}{2} < \omega \leq \pi]$) and then projected into a Ramachandran projection discretized over a 18$\times$18 $\phi$-$\psi$ grid. The high computational cost of these calculations means that it was computationally intractable to conduct \textit{ab initio} energy evaluations for all snapshots. Instead, for each $\phi$-$\psi$ grid cell populated by trajectory snapshots, we elect to compute and compare the \textit{ab initio} potential energy for the frame with the lowest classical potential energy in each of the cis and trans $\omega$ states. Uncertainties were estimated by block averaging by splitting the classical trajectory into 5$\times$100 ns fragments and repeating this procedure for each block. Single-point potential energy calculations were performed using second-order M{\o}ller-Plesset perturbation theory (MP2) \cite{moller1934note} with a 6-31G** basis set \cite{ditchfield1971self,hariharan1973influence,hehre1972self} as implemented in PySCF \cite{pyscf_1,pyscf_2} (Version 2.4.0), which utilizes \texttt{libcint} \cite{libcint} (Version 6.0.0) for integral evaluations. The single-point energies for the Nipe and Nipm monomers with iodine-containing side chains were calculated at MP2/6-311G** level of theory \cite{moller1934note,krishnan1980self}. 

\section{Results}

We test the predictive accuracy of the MoSiC-CGenFF-NTOID model in three ways. First, we compare potential energy distributions over Ramachandran plots against \textit{ab initio} calculations. Second, we compare cis/trans isomerization energies and free energies against \textit{ab initio} calculations and experimental measurements. Third, we compare the predictions of our model for peptoid self-assembly to experimental observations \cite{Castelletto2020}.

\subsection{Peptoid Ramachandran Plot Calculations}\label{subsec:rama}

Our first test was to assess the ability of the MoSiC-CGenFF-NTOID model to accurately recapitulate the \textit{ab initio} potential energies. To do so, we draw a comparison between potential energy landscapes of end-capped peptoid monomers projected onto Ramachandran projections into the $\phi$-$\psi$ backbone dihedrals partitioned by the cis and trans states of the $\omega$ dihedral (Section \ref{subsec:energy}). Effectively, the constitutes a comparison of the consistency of the quantum and classical potential energies as a function of the $\phi$, $\psi$, and $\omega$ dihedral angle states. This type of comparison has precedent as a means to evaluate peptoid all-atom force fields \cite{https://doi.org/10.1002/jcc.25850, Harris2023, mukherjee2015, Hoyas2018, https://doi.org/10.1002/jcc.23478, Du2018}. Good agreement between the (relative) potential energies $U(\phi,\psi; \omega = \text{cis/trans})$ would provide support for the accuracy of the MoSiC-CGenFF-NTOID model in reproducing the molecular energy landscape across a variety of peptoid side chains.

\begin{figure*}[ht!]
    \centering
    \includegraphics[width=\textwidth]{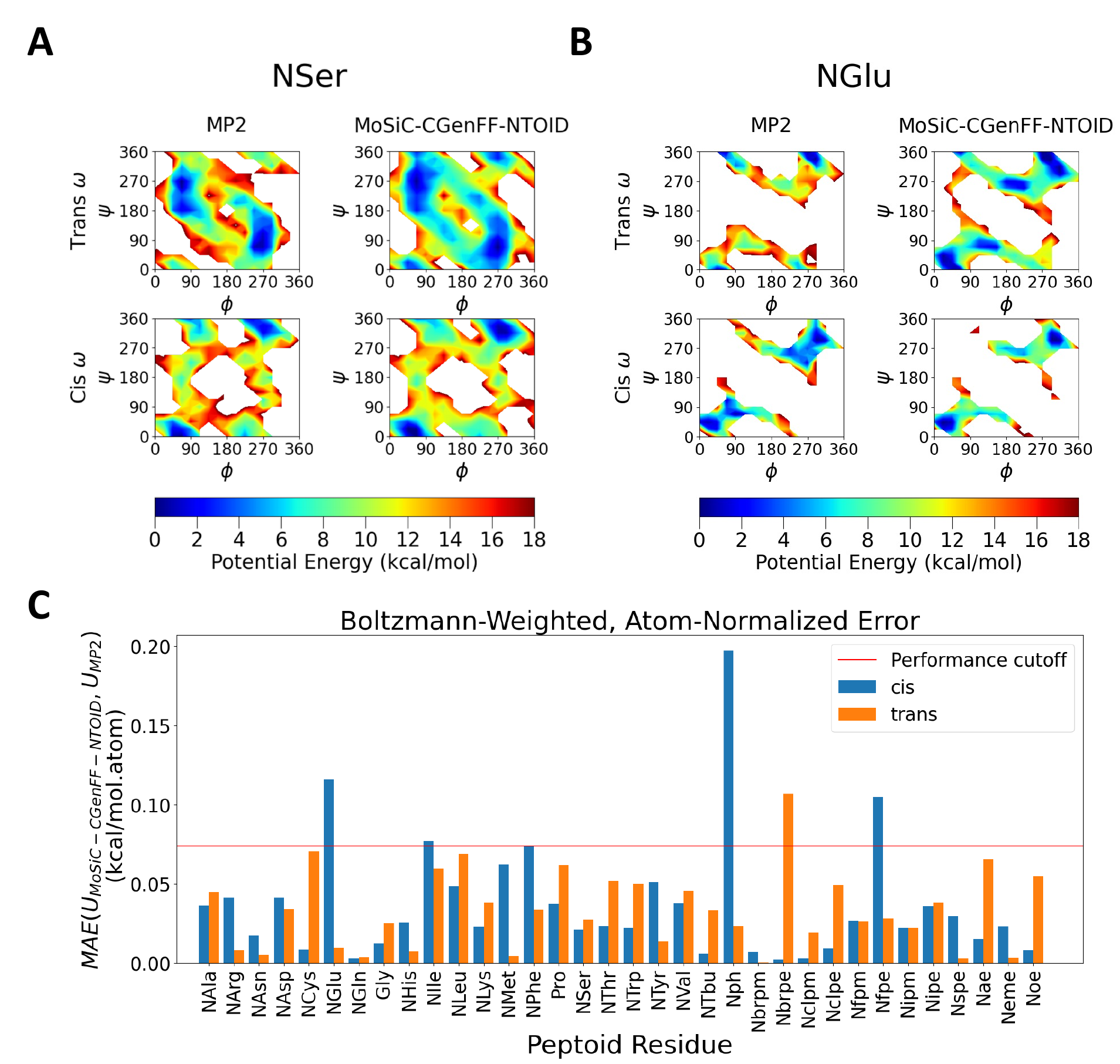}
    \caption{Comparison of classical MoSiC-CGenFF-NTOID and \textit{ab initio} MP2 potential energies in vacuum projected onto Ramachandran plot coordinates in the $\phi$ and $\psi$ backbone dihedrals for acetyl N-capped and aminomethyl C-capped peptoid monomers. Comparisons of the potential energy landscapes $U(\phi,\psi)$ in the cis and trans $\omega$ states for (A) N-Serine (NSer) and (B) N-Glutamate (NGlu) side chains are presented as illustrative examples of good and poor performing MoSiC-CGenFF-NTOID parameterizations. Analogous plots for the remaining 32 peptoid side chains considered are presented in \blauw{Figures S7-S38}. The upper row of each quartet illustrates the trans configurations ($[-\pi < \omega \leq -\frac{\pi}{2}] \cup [\frac{\pi}{2} < \omega \leq \pi]$) and the lower row the cis ($-\frac{\pi}{2} < \omega \leq \frac{\pi}{2}$). The left column illustrates the \textit{ab initio} MP2 potential energies and the right column the MoSiC-CGenFF-NTOID potential energies. Since only relative energies are relevant to thermodynamic properties, the zero of potential energy for visualization purposes is set to the global energy minimum. (C) Atom-count normalized, Boltzmann-weighted mean absolute errors between the classical MoSiC-CGenFF-NTOID and \textit{ab initio} MP2 Ramachandran potential energy landscapes averaged over all frames for which comparisons were made.}
    \label{fig:rama}
\end{figure*}

We present in Figure \ref{fig:rama} comparisons of $U(\phi,\psi; \omega = \text{cis/trans})$ for the N-Serine (NSer) and N-Glutamate (NGlu) side chains as illustrative examples of good and poor performing comparisons.  Analogous plots for the remaining 32 residues are presented in \blauw{Figures S7-S38}. For the 29/34 residues exhibiting good agreement in the Ramachandran potential energy landscapes, of which NSer is a representative example (Figure \ref{fig:rama}A), we observe both the location and depth of the minima and maxima to be in good agreement between the MoSiC-CGenFF-NTOID and MP2 landscapes. For the 5/34 residues exhibiting poor agreement, of which NGlu is a representative example (Figure \ref{fig:rama}B), we qualitatively observe that the primary source of discrepancy tends to be a smoother energy landscape and broader stable regions exhibited by classical energy calculations compared to the quantum calculations. While the locations of the potential energy minima within the Ramachandran projections tend to be in good agreement, the depth of the free energy wells differ. 

We quantify the degree of agreement in the classical and \textit{ab initio} potential energy landscapes by computing the atom-count normalized, Boltzmann-weighted mean absolute error averaged over all frames for which comparisons were made,
\begin{widetext}
\begin{align}\label{eq:mae_rama}
 MAE(U_{MoSiC-CGenFF-NTOID},U_{MP2}) 
 = \frac{1}{N_A N_S}\frac{\sum_i \; \left|U_{MoSiC-CGenFF-NTOID}(\phi_i, \psi_i) - U_{MP2}(\phi_i, \psi_i)\right| e^{-U_{MP2}(\phi_i, \psi_i)/k_BT}}{\sum_i \; e^{-U_{MP2}(\phi_i, \psi_i)/k_BT}} 
\end{align}
\end{widetext}
where $N_A$ is the number of atoms in the molecule, $i = 1 \ldots N_S$ indexes the structures for which energies were calculated, $U_{MoSiC-CGenFF-NTOID}(\phi_i, \psi_i)$ and $U_{MP2}(\phi_i, \psi_i)$ are the potential energies calculated with our force field and quantum-mechanical methods respectively, $k_B$ is Boltzmann's constant, and $T$ is the absolute temperature. We report these comparisons in Table \ref{tab:master} and illustrate them visually in Figure \ref{fig:rama}C. The corresponding non-Boltzmann weighted and non-atom-count normalized plots are presented in \blauw{Figure S2}. 

As mentioned above, the parameterization of three side chains -- methyl (NAla), s-(1)-phenylethyl (Nspe), and phenylmethyl (NPhe) -- within MoSiC-CGenFF-NTOID incorporated a refitted $\chi_1$ dihedral parameterized against \textit{ab initio} calculations by Weiser and Santiso \cite{https://doi.org/10.1002/jcc.25850}. Of these, the  poorest performing MAE resulted from the cis state of the NPhe residue, possessing a $MAE(U_{MoSiC-CGenFF-NTOID},U_{MP2})$ = 0.074 kcal/mol.atom. We adopted this threshold as the cutoff by which to quantify good versus poor agreement of the Ramachandran potential energy landscapes. Side chains deemed to possess good performance under this metric have both cis (blue) and trans (orange) bars in Figure \ref{fig:rama}C lying below the horizontal red line and the corresponding MAE values are colored in green in the second column of Table \ref{tab:master}. The MAE values of the poor performing side chains are colored red and number only five -- NGlu, NIle, Nph, Nbrpe, and Nfpe. 


\begin{table*}
    \centering
        
    \caption{Performance assessment of MoSiC-CGenFF-NTOID in comparison to MP2 \textit{ab initio} energy calculations for each of the 34 peptoid side chains. We report the atom-count normalized, Boltzmann-weighted mean absolute error (MAE) over the Ramachandran $\phi$-$\psi$ potential energy landscape (Eqn.\ \ref{eq:mae_rama}) in the second column, and the cis/trans energy difference $\Delta U_{\text{cis/trans}}$ computed under MoSiC-CGenFF-NTOID and its absolute error relative to MP2 (Eqn.\ \ref{eq:mae_ct}) in the third column. The threshold for good performance is set based on the performance of the NPhe side chain explicitly reparameterized agains \textit{ab initio} calculations by Weiser and Santiso \cite{https://doi.org/10.1002/jcc.25850}, under which we define a cutoff of MAE = 0.074 kcal/mol.atom for the Ramachandran MAE and 8.04 kcal/mol for both the $\Delta U_{\text{cis/trans}}$ discrepancy and uncertainty in this value. Side chains which pass these thresholds have their values reported in green. Side chains which fail to pass these thresholds have their values reported in red and bold-faced red text identifies the MAE, $Err$, or uncertainty value that causes the performance to be classified as poor. Side chains that pass both assessments are deemed as good, those which fail one of the two as fair, and those which fail both as poor, as indicated in the fourth column.}
    \begin{adjustbox}{width=\textwidth,center=\textwidth}
    \begin{tabular}{|l|cc|cc|c|}
    \toprule

Residue &  \multicolumn{2}{c|}{Ramachandran MAE (kcal/mol.atom) (Eqn.\ \ref{eq:mae_rama})} & \multicolumn{2}{c|}{$\Delta U_{\text{cis/trans}}$ (kcal/mol)}  & Recommendation \\
&  Cis & Trans & MoSiC-CGenFF-NTOID Value & Error (Eqn.\ \ref{eq:mae_ct})  & \\
\hline
NAla &  \textcolor{OliveGreen}{0.036 } & \textcolor{OliveGreen}{0.045 } & \textcolor{OliveGreen}{11.51 $\pm$ 1.52 } & \textcolor{OliveGreen}{0.21 } &\textcolor{OliveGreen}{Good} \\ 
NArg &  \textcolor{OliveGreen}{0.041 } & \textcolor{OliveGreen}{0.008 } & \textcolor{OliveGreen}{-3.14 $\pm$ 1.02 } & \textcolor{OliveGreen}{0.18 } &\textcolor{OliveGreen}{Good} \\ 
NAsn &  \textcolor{OliveGreen}{0.018 } & \textcolor{OliveGreen}{0.005 } & \textcolor{OliveGreen}{-21.20 $\pm$ 5.78 } & \textcolor{OliveGreen}{2.16 } &\textcolor{OliveGreen}{Good} \\ 
NAsp &  \textcolor{OliveGreen}{0.041 } & \textcolor{OliveGreen}{0.034 } & \textcolor{OliveGreen}{9.30 $\pm$ 2.18 } & \textcolor{OliveGreen}{3.79 } &\textcolor{OliveGreen}{Good} \\ 
NCys &  \textcolor{OliveGreen}{0.008 } & \textcolor{OliveGreen}{0.071 } & \textcolor{OliveGreen}{7.22 $\pm$ 3.79 } & \textcolor{OliveGreen}{2.63 } &\textcolor{OliveGreen}{Good} \\ 
NGlu &  \textcolor{red}{\textbf{0.116} } & \textcolor{red}{0.010 } & \textcolor{OliveGreen}{4.55 $\pm$ 6.48 } & \textcolor{OliveGreen}{5.49 } &\textcolor{orange}{Fair} \\ 
NGln &  \textcolor{OliveGreen}{0.003 } & \textcolor{OliveGreen}{0.004 } & \textcolor{OliveGreen}{-7.11 $\pm$ 4.18 } & \textcolor{OliveGreen}{4.90 } &\textcolor{OliveGreen}{Good} \\ 
Gly &  \textcolor{OliveGreen}{0.012 } & \textcolor{OliveGreen}{0.025 } & \textcolor{OliveGreen}{14.28 $\pm$ 2.61 } & \textcolor{OliveGreen}{5.07 } &\textcolor{OliveGreen}{Good} \\ 
NHis &  \textcolor{OliveGreen}{0.026 } & \textcolor{OliveGreen}{0.007 } & \textcolor{OliveGreen}{22.87 $\pm$ 2.79 } & \textcolor{OliveGreen}{2.44 } &\textcolor{OliveGreen}{Good} \\ 
NIle &  \textcolor{red}{\textbf{0.077} } & \textcolor{red}{0.060 } & \textcolor{OliveGreen}{15.02 $\pm$ 5.32 } & \textcolor{OliveGreen}{2.47 } &\textcolor{orange}{Fair} \\ 
NLeu &  \textcolor{OliveGreen}{0.048 } & \textcolor{OliveGreen}{0.069 } & \textcolor{red}{29.52 $\pm$ \textbf{8.38} } & \textcolor{red}{1.05 } &\textcolor{orange}{Fair} \\ 
NLys &  \textcolor{OliveGreen}{0.023 } & \textcolor{OliveGreen}{0.038 } & \textcolor{OliveGreen}{-4.53 $\pm$ 2.94 } & \textcolor{OliveGreen}{6.16 } &\textcolor{OliveGreen}{Good} \\ 
NMet &  \textcolor{OliveGreen}{0.062 } & \textcolor{OliveGreen}{0.004 } & \textcolor{OliveGreen}{15.19 $\pm$ 4.40 } & \textcolor{OliveGreen}{1.65 } &\textcolor{OliveGreen}{Good} \\ 
NPhe &  \textcolor{OliveGreen}{0.074 } & \textcolor{OliveGreen}{0.034 } & \textcolor{OliveGreen}{5.55 $\pm$ 2.62 } & \textcolor{OliveGreen}{4.02 } &\textcolor{OliveGreen}{Good} \\ 
Pro &  \textcolor{OliveGreen}{0.037 } & \textcolor{OliveGreen}{0.062 } & \textcolor{OliveGreen}{-6.58 $\pm$ 3.55 } & \textcolor{OliveGreen}{2.55 } &\textcolor{OliveGreen}{Good} \\ 
NSer &  \textcolor{OliveGreen}{0.021 } & \textcolor{OliveGreen}{0.028 } & \textcolor{OliveGreen}{9.84 $\pm$ 4.87 } & \textcolor{OliveGreen}{3.41 } &\textcolor{OliveGreen}{Good} \\ 
NThr &  \textcolor{OliveGreen}{0.023 } & \textcolor{OliveGreen}{0.052 } & \textcolor{OliveGreen}{-13.64 $\pm$ 5.28 } & \textcolor{OliveGreen}{1.37 } &\textcolor{OliveGreen}{Good} \\ 
NTrp &  \textcolor{OliveGreen}{0.022 } & \textcolor{OliveGreen}{0.050 } & \textcolor{red}{-7.61 $\pm$ \textbf{18.84} } & \textcolor{red}{\textbf{9.77} } &\textcolor{orange}{Fair} \\ 
NTyr &  \textcolor{OliveGreen}{0.051 } & \textcolor{OliveGreen}{0.014 } & \textcolor{OliveGreen}{9.07 $\pm$ 3.21 } & \textcolor{OliveGreen}{0.36 } &\textcolor{OliveGreen}{Good} \\ 
NVal &  \textcolor{OliveGreen}{0.038 } & \textcolor{OliveGreen}{0.045 } & \textcolor{OliveGreen}{11.03 $\pm$ 6.70 } & \textcolor{OliveGreen}{1.13 } &\textcolor{OliveGreen}{Good} \\ 
Ntbu &  \textcolor{OliveGreen}{0.006 } & \textcolor{OliveGreen}{0.033 } & \textcolor{OliveGreen}{-6.53 $\pm$ 4.15 } & \textcolor{OliveGreen}{5.00 } &\textcolor{OliveGreen}{Good} \\ 
Nph &  \textcolor{red}{\textbf{0.197} } & \textcolor{red}{0.024 } & \textcolor{red}{5.90 $\pm$ 4.39 } & \textcolor{red}{\textbf{10.96} } &\textcolor{red}{Poor} \\ 
Nbrpm &  \textcolor{OliveGreen}{0.007 } & \textcolor{OliveGreen}{0.000 } & \textcolor{OliveGreen}{-0.00 $\pm$ 3.37 } & \textcolor{OliveGreen}{1.31 } &\textcolor{OliveGreen}{Good} \\ 
Nbrpe &  \textcolor{red}{0.002 } & \textcolor{red}{\textbf{0.107} } & \textcolor{OliveGreen}{16.34 $\pm$ 5.57 } & \textcolor{OliveGreen}{1.46 } &\textcolor{orange}{Fair} \\ 
Nclpm &  \textcolor{OliveGreen}{0.003 } & \textcolor{OliveGreen}{0.019 } & \textcolor{OliveGreen}{12.80 $\pm$ 3.04 } & \textcolor{OliveGreen}{2.30 } &\textcolor{OliveGreen}{Good} \\ 
Nclpe &  \textcolor{OliveGreen}{0.009 } & \textcolor{OliveGreen}{0.049 } & \textcolor{OliveGreen}{11.31 $\pm$ 7.41 } & \textcolor{OliveGreen}{2.15 } &\textcolor{OliveGreen}{Good} \\ 
Nfpm &  \textcolor{OliveGreen}{0.027 } & \textcolor{OliveGreen}{0.026 } & \textcolor{OliveGreen}{10.40 $\pm$ 3.19 } & \textcolor{OliveGreen}{1.60 } &\textcolor{OliveGreen}{Good} \\ 
Nfpe &  \textcolor{red}{\textbf{0.105} } & \textcolor{red}{0.028 } & \textcolor{OliveGreen}{15.59 $\pm$ 2.46 } & \textcolor{OliveGreen}{2.79 } &\textcolor{orange}{Fair} \\ 
Nipm &  \textcolor{OliveGreen}{0.022 } & \textcolor{OliveGreen}{0.022 } & \textcolor{OliveGreen}{10.14 $\pm$ 3.48 } & \textcolor{OliveGreen}{1.16 } &\textcolor{OliveGreen}{Good} \\ 
Nipe &  \textcolor{OliveGreen}{0.036 } & \textcolor{OliveGreen}{0.038 } & \textcolor{OliveGreen}{11.88 $\pm$ 3.49 } & \textcolor{OliveGreen}{3.73 } &\textcolor{OliveGreen}{Good} \\ 
Nspe &  \textcolor{OliveGreen}{0.029 } & \textcolor{OliveGreen}{0.003 } & \textcolor{OliveGreen}{-4.67 $\pm$ 3.57 } & \textcolor{OliveGreen}{2.07 } &\textcolor{OliveGreen}{Good} \\ 
Nae &  \textcolor{OliveGreen}{0.015 } & \textcolor{OliveGreen}{0.066 } & \textcolor{red}{-10.61 $\pm$ 3.36 } & \textcolor{red}{\textbf{12.83} } &\textcolor{orange}{Fair} \\ 
Neme &  \textcolor{OliveGreen}{0.023 } & \textcolor{OliveGreen}{0.003 } & \textcolor{OliveGreen}{-10.01 $\pm$ 4.26 } & \textcolor{OliveGreen}{2.93 } &\textcolor{OliveGreen}{Good} \\ 
Noe &  \textcolor{OliveGreen}{0.008 } & \textcolor{OliveGreen}{0.055 } & \textcolor{red}{19.16 $\pm$ \textbf{10.97} } & \textcolor{red}{\textbf{8.15} } &\textcolor{orange}{Fair} \\ 
\hline
    \end{tabular}
    \end{adjustbox}
   
    \label{tab:master}
\end{table*}

\subsection{Cis-Trans Equilibrium of Individual Residues and Polymers}


In contrast to peptides, which almost exclusively exist in a trans amide bond state, the tertiary amide bond within peptoids permits the $\omega$ dihedral to access both the cis and trans configurations \cite{Weiser2017, https://doi.org/10.1002/jcc.23478, Voelz2010}. The activation barrier separating these two states is sequence dependent, but is typically on the order of $\sim$15 kcal/mol \cite{Weiser2017, feigel1983rotation, Qiang2007}. Sampling these $\omega$ dihedral isomerizations remains challenging in molecular simulations due to the comparatively high energy barrier and associated slow time scales relative to other peptoid conformational changes \cite{Zhao2020}, but is of critical importance in understanding and engineering peptoids with desired self-assembly patterns \cite{Eastwood2023} and biological functions\cite{Qiu2023}. The choice of side chain has a profound effect on the cis/trans equilibrium \cite{Kalita2022, Eastwood2023, doi:10.1073/pnas.1800397115, Roy2013} and thus is an important benchmark to be addressed by a force field designed to simulate a diversity of side chains.
 
Our second test assessed our force field's ability to simulate cis/trans equilibria in three separate evaluations. First, we compare cis-trans potential energy differences computed through our MoSiC-CGenFF-NTOID package to those calculated through quantum methods. Second, we evaluate whether the force field captures the expected cis/trans preferences induced by different side chains by comparing free energy differences $\Delta G_{\text{cis/trans}}$. Third, we compare the calculated cis/trans free energy differences with the experimentally reported values available for a small number of peptoids. 

\subsubsection{Cis-Trans Potential Energy Difference Calculations in Monomers}

Our first cis/trans evaluation assesses the capacity of the MoSiC-CGenFF-NTOID force field to accurately recapitulate peptoid cis/trans isomerization energy differences. We adopt a similar test to that in Section \ref{subsec:rama}, but instead of analyzing the entire Ramachandran potential energy plot, we focus on the energy difference $\Delta U_{\text{cis/trans}}$ between the cis and trans states. Good agreement between values will indicate that our force field reliably predicts the energetics of this isomerization as a critical prerequisite for the accurate prediction of peptoid structure and dynamics.

We conduct this comparison by comparing the cis/trans energy difference computed using MoSiC-CGenFF-NTOID to that calculated using MP2 \textit{ab initio} calculations. We quantify the agreement via the absolute error,
\begin{align}\label{eq:mae_ct}
    Err(\Delta U^{MoSiC-CGenFF-NTOID}_{\text{cis/trans}}, \: \Delta U^{MP2}_{\text{cis/trans}}) \\\nonumber = \left|\overline{\Delta U}^{MP2}_{\text{cis/trans}} -  \overline{\Delta U}^{MoSiC-CGenFF-NTOID}_{\text{cis/trans}}\right|
\end{align}
where $\overline{\Delta U}_{\text{cis/trans}} = \overline{U^*_{\text{cis}} - U^*_{\text{trans}}}$, $U^*_\text{cis/trans}$ denotes the potential energy of the most probable $\phi$-$\psi$ configuration in either the cis or trans state, and the overbar denotes an average is taken over five frames, one from each of five independent 100 ns trajectories.

\begin{figure*}
    \centering
    \includegraphics[width=\textwidth]{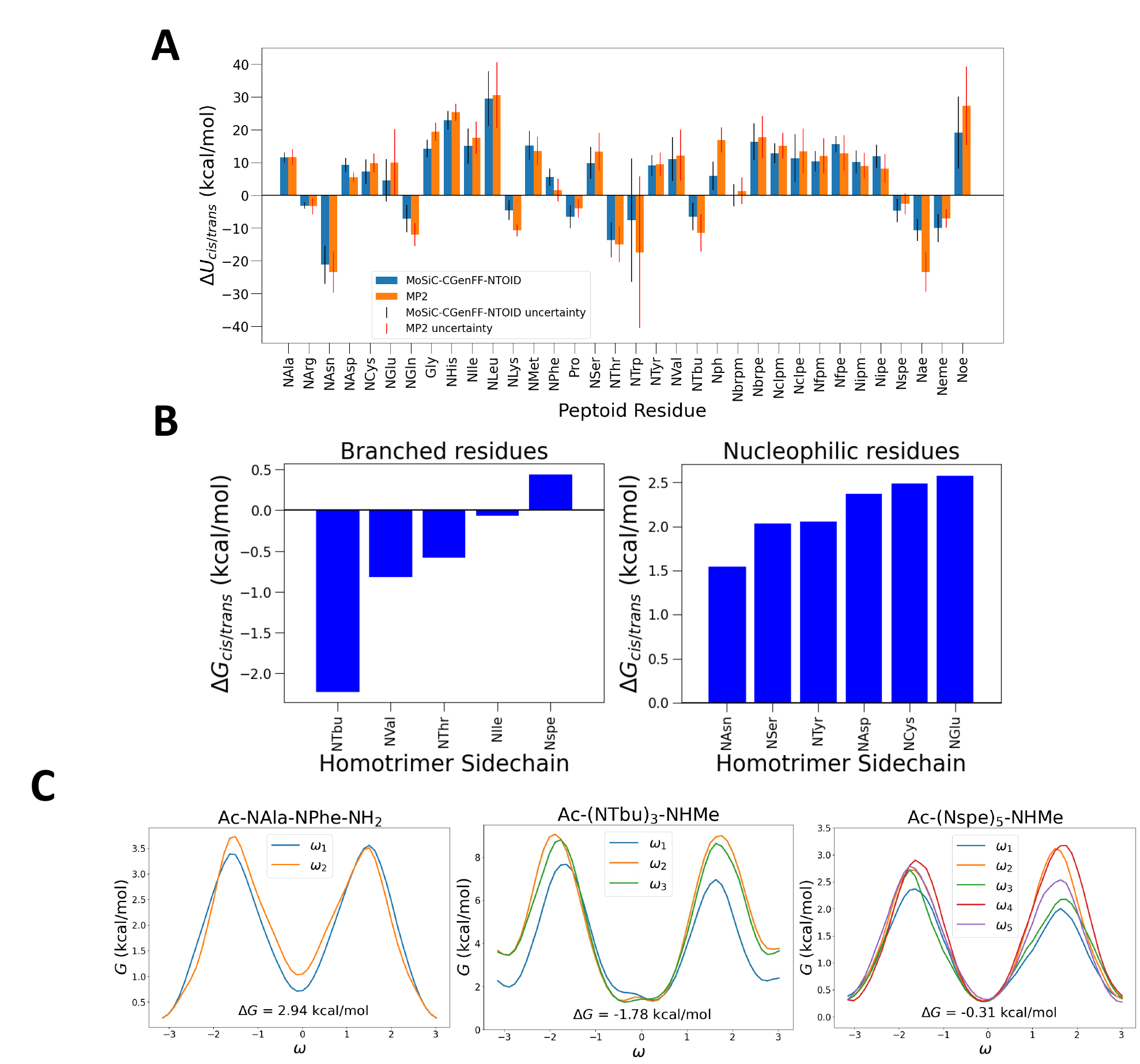}
    \caption{Comparison of MoSiC-CGenFF-NTOID predictions of cis/trans potential and free energy differences compared to \textit{ab initio} calculations and experimental measurements. (A) The potential energy difference $\Delta U_{\text{cis/trans}}$ between the cis and trans states from MoSiC-CGenFF-NTOID and \textit{ab initio} MP2 calculations. (B) $\Delta G_{\text{cis/trans}} = G_\text{cis} - G_\text{trans}$ calculated under the MoSiC-CGenFF-NTOID force field over the central $\omega_2$ dihedral for peptoid homotrimers possessing branched (left) and nucleophilic (right) side chains. As anticipated, the branched side chains tend to favor cis isomeric states, whereas the nucleophilic side chains tend to favor trans isomeric states. (C) Free energy profiles in each $\omega_i$ dihedral within the three peptoid sequences -- Ac-NAla-NPhe-NH\textsubscript{2}, Ac-(NTbu)\textsubscript{3}-NHMe, and Ac-(Nspe)\textsubscript{5}-NHMe -- for which experimental $\Delta G_{\text{cis/trans}}$ values are available.}
    \label{fig:cistrans}
\end{figure*}

As in Section \ref{subsec:rama}, we use NPhe, the parametrization of which is unchanged in MoSiC-CGenFF-NTOID from the Weiser and Santiso \textit{ab intio} refitting, as a baseline for performance assessment, which possesses a $Err(\Delta U^{MoSiC-CGenFF-NTOID}_{\text{cis/trans}}, \: \Delta U^{MP2}_{\text{cis/trans}})$ = 4.02 kcal/mol. We deemed a side chain to exhibit acceptable performance if it has both an $Err$ value and an uncertainty among the five calculated $\Delta U^{MoSiC-CGenFF-NTOID}_{\text{cis/trans}}$ values within a threshold of twice this value (i.e., 8.04 kcal/mol). We report the assessment of our 34 side chain residues under this criterion in Figure \ref{fig:cistrans}A and Table \ref{tab:master}. Five side chains -- NLeu, NTrp, Nph, Nae, and Noe -- were deemed unacceptable under this criterion, and are highlighted in red text within the third column of Table \ref{tab:master}.

The performance under both the Ramachandran and $\Delta U_{\text{cis/trans}}$ criteria constitute our two quantitative assessments of performance available for all 34 side chain, we deem a side chain to possess good performance if it passes both assessments, fair performance if it passes only one of the two, and poor performance if it fails both. Under this definition, 25/34 side chains are good, 8/34 fair, and only 1/34 poor. Notably, the only unacceptable performer is Nph. Only three side chains -- Gly, Pro, and Nph -- do not belong to the family of substituted methyl groups. Crucially, the CGenFF-NTOID peptoid backbone parameterization developed by Weiser and Santiso via reparameterization of the $\rho$, $\psi$, and $\omega$ backbone dihedrals, was explicitly designed to pertain to side chains connected via methylene or substituted methylene bridges (i.e., substituted methyl groups \ce{-CH_3}, \ce{-CH_2R}, \ce{-CHRR$^{\prime}$} \ce{-CRR$^{\prime}$R$^{\prime\prime}$}) \cite{https://doi.org/10.1002/jcc.25850}. The poor Nph performance reinforces the guidance that explicit reparameterization of the $\rho$ and $\chi_1$ dihedrals is likely required for side chains not falling into this family. The comparatively good performance of Gly and Pro can be understood as special cases of peptoid residues that are identical to their peptide analogs and for which we adopted the existing peptide parameterization within CGenFF. In light of these observations, we propose the Ramachandran potential energy and $\Delta U_{\text{cis/trans}}$ comparisons as a means to assess whether reparameterization of a side chain may be warranted.

\subsubsection{Recapitulation of Known Trends in Cis/Trans Equilibria}\label{subsec:homopolymer_dgct}
Our second cis/trans evaluation assesses the degree to which residues possessing expected cis-promoting and trans-promoting moieties exhibit these properties within the MoSiC-CGenFF-NTOID force field. Specifically, branched side chains are expected to favor cis states due to the relatively smaller oxygen being more accommodating to the steric bulk of a branched side chain in a cis state than the $\alpha$-carbon and $\alpha$-hydrogens of the adjacent methylene group  \cite{Gorske2009-it}, whereas nucleophilic residues, which disrupt n--$\pi^*$ interactions between backbone and side chain atoms and promote n--$\pi^*$ interactions between backbone atoms, are expected to favor trans states\cite{Knight2015-il}. Thermodynamically, this leads to the expectation that peptoid homopolymers comprising branched residues should possesses lower (more negative) $\Delta G_{\text{cis/trans}} = G_\text{cis} - G_\text{trans}$ values than nucleophilic residues that should possess higher (more positive) values. We constructed homotrimers of five branched peptoid side chains -- NTbu, NVal, NThr, NIle, and Nspe -- and six nucleophilic side chains -- NAsn, NSer, NTyr, NAsp, NCys, and NGlu -- and calculated $\Delta G_{\text{cis/trans}}$ by reweighting the WT-PBMetaD enhanced sampling results into the unbiased ensemble, calculating populations of  the cis ($-\frac{\pi}{2} < \omega \leq \frac{\pi}{2}$) and trans ($[-\pi < \omega \leq -\frac{\pi}{2}] \cup [\frac{\pi}{2} < \omega \leq \pi]$) states, and extracting the resulting free energy difference. We report the results of these calculations in Figure \ref{fig:cistrans}B, and $\omega_2$ free energy surfaces for homotrimers of all 34 peptoid side chains considered in this work are presented in \blauw{Figure S3}. As anticipated, the branched residues to tend to possess negative or weakly positive $\Delta G_{\text{cis/trans}}$ values that tend to favor cis isomers, whereas the nucleophilic residues possess strongly positive $\Delta G_{\text{cis/trans}}$ values that promote the trans isomer. 
These comparisons provide support that the MoSiC-CGenFF-NTOID force field recapitulates cis/trans thermodynamic preferences in line with expectations based on the physicochemical character of the peptoid side chains.

\subsubsection{Comparison of Experimental Cis/Trans Equilibrium Constants in MoSiC-CGenFF-NTOID and NMR Spectroscopy}
Our third cis/trans evaluation compares $\Delta G_{\text{cis/trans}}$ values derived from our simulations with those measured in experiment. Experimental measurements of cis-trans equilibrium have been conducted for a large number of peptoids, but the majority correspond to those with side chains, terminal caps, or solution conditions currently unsupported by our model. However, we have identified three sequences for which nuclear magnetic resonance (NMR) spectroscopy measurements of $\Delta G_{\text{cis/trans}}$ are available for comparison with MoSiC-CGenFF-NTOID simulations of Ac-NAla-NPhe-NH\textsubscript{2}, Ac-(NTbu)\textsubscript{3}-NHMe, and Ac-(Nspe)\textsubscript{5}-NHMe \cite{Qiang2007, Roy2013, Stringer2011-us}.  In Figure \ref{fig:cistrans}C, we report the free energy profiles computed in each $\omega$ dihedral from our MoSiC-CGenFF-NTOID simulations, and, for the purposes of experimental comparison, report in Table \ref{tab:exptkct} simple averages over the $\Delta G_{\text{cis/trans}}$ for each dihedral. 
\begin{table}[h]
  \caption{Comparison between $\Delta G_{\text{cis/trans}}$ values calculated by MoSiC-CGenFF-NTOID and experimentally measured by NMR spectroscopy. Simulations were conducted in water at 300 K and 1 bar, employ acetyl (Ac) N-caps and amino (NH\textsubscript{2}) or aminomethyl (NHMe) C-caps, and standard errors in the computational values estimated by 4-fold block averaging. Experimental measurements were conducted at standard temperature and pressure and in variety of solvents as indicated by footnotes to the table.} 
  \begin{ruledtabular}\label{tab:exptkct}

    \begin{tabular}{lp{25mm}p{25mm}}
    Peptoid Sequence & NMR $\Delta G_{\text{cis/trans}}$ \newline (kcal/mol) & MoSiC-CGenFF-NTOID $\Delta G_{\text{cis/trans}}$ \newline (kcal/mol) \\
    \hline
    Ac-NAla-NPhe-NH$_2$\cite{Qiang2007}\footnote{Water solvent.}& 0.48 & 2.95 $\pm$ 0.06 \\
    Ac-(NTbu)$_3$-NHMe\cite{Roy2013}\footnote{Deuterated hexane, acetonitrile, methanol, acetone, and dimethyl sulfoxide (DMSO) solvents.} & $<$-1.74 & -1.78 $\pm$ 0.06 \\
    Ac-(Nspe)$_5$-NHMe\cite{Stringer2011-us}\textsuperscript{,}\footnote{Acetonitrile solvent. The experimental sequence employed a tert-butyoxy (OtBu) rather than NHMe C-cap, which is expected to promote elevated cis character (i.e., more negative $\Delta G_{\text{cis/trans}}$) due to additional steric strain between the the OtBu C-cap and the adjacent methylene group.} & -0.71 & -0.31 $\pm$ 0.25 \\
    \end{tabular}
\end{ruledtabular}
\end{table}

 The signs of the MoSiC-CGenFF-NTOID $\Delta G_{\text{cis/trans}}$ values match those of the experimental measurements in all three instances, providing good qualitative support for the capacity of the model to predict the overall cis/trans preference in these sequences. There are, however, quantitative discrepancies in the range 0.04-2.5 kcal/mol. The $\Delta G_{\text{cis/trans}}$ values are known to be quite sensitive to subtle changes in the molecular chemistry and solvent environment. For example, $\Delta G_{\text{cis/trans}}$ values between -0.71 and -0.10 kcal/mol have been reported for Nspe depending on the degree of polymerization, identity of the terminal cap, and the solution conditions \cite{Stringer2011-us, Gorske2009-it, Gimenez2019-ds}. To explore the sensitivity of $\Delta G_{\text{cis/trans}}$ within the model and quantify the magnitude of expected discrepancies based on different choices of these parameters, we conducted two suites of additional simulations to explore the influence of residue ordering and choice of C-terminal cap. First, we conducted WT-PBMetaD simulations of NPhe--Nae--NPhe (FXF) and Nae--NPhe--NPhe (XFF) heterotrimers with acetyl (Ac) N-caps and aminomethyl (NHMe) C-caps, and computed and report in in Table \ref{tab:conditionkct} the $\Delta G_{\text{cis/trans}}$ for the $\omega_2$ and $\omega_3$ dihedrals. While the cis/trans preference for the $\omega_3$ dihedral is essentially unchanged as a function of sequence, for the $\omega_2$ dihedral we observe a $\Delta G_{\text{cis/trans}}$ difference of 1.60 kcal/mol, corresponding to a $\sim$15-fold increase in the preference of cis states in FXF relative to XFF as measured by the equilibrium constant $K_{\text{cis/trans}} = \exp{(-\Delta G_{\text{cis/trans}} / k_B T)}$. Second, we conducted WT-PBMetaD simulations of (Nspe)$\textsubscript{3}$ (ZZZ) with an acetyl (Ac) N-cap and three different C-caps: aminomethyl (NHMe), aminodimethyl (N(Me)\textsubscript{2}), and amino (NH\textsubscript{2}). Again, we observe the $\Delta G_{\text{cis/trans}}$ of the $\omega_3$ dihedral not to change by a large margin, remaining within a range of 0.35 kcal/mol regardless of the C-cap. Conversely, the $\Delta G_{\text{cis/trans}}$ of the $\omega_2$ dihedral, which is located closer to the N-terminal chain end varies by 0.84 kcal/mol between --NH$_2$-capped and --NHMe-capped peptoids, corresponding a $\sim$4-fold increase in the preference for the cis-state in the former. The relatively high variability in $\Delta G_{\text{cis/trans}}$ provides motivation for additional experimental measurements and subsequent force field validation, and if necessary reparameterization, to assure predictions are robust and transferable to the range of possible peptoid sequences, end caps, and solvent conditions. Within the scope of the present work, the qualitative agreement of the simulation and experimental $\Delta G_{\text{cis/trans}}$ values in the face of these uncertainties, can be viewed as good additional support for the predictive accuracy of MoSiC-CGenFF-NTOID.

\begin{table}
\caption{Dependence of MoSiC-CGenFF-NTOID predicted cis/trans free energy differences $\Delta G_{\text{cis/trans}}$ as a function of residue ordering and choice of C-terminal capping group. }
\begin{ruledtabular}
     \centering
    \begin{tabular}{lcccr}\label{tab:conditionkct}
      Sequence & N-cap & C-cap & Dihedral & $\Delta G_{\text{cis/trans}}$ \\
      &  &  & & (kcal/mol) \\
      \hline
      FXF & Ac  & NHMe & $\omega_2$ & -0.52 $\pm$ 0.12 \\
      FXF& Ac &  NHMe  &$\omega_3$ & 1.61 $\pm$ 0.40 \\
      XFF & Ac  & NHMe& $\omega_2$ & 1.08 $\pm$ 0.07\\
      XFF & Ac & NHMe & $\omega_3$ & 1.59 $\pm$ 0.30 \\
      \hline
      ZZZ  & Ac & N(Me)$_2$ & $\omega_2$ & -0.16 $\pm$ 0.14 \\
      ZZZ & Ac&  N(Me)$_2$ &$\omega_3$& 0.07 $\pm$ 0.09 \\
      ZZZ & Ac  & NHMe & $\omega_2$ & 0.49 $\pm$ 0.28 \\
      ZZZ & Ac &  NHMe &$\omega_3$ & 0.42 $\pm$ 0.28 \\
      ZZZ & Ac  & NH$_2$ & $\omega_2$ & -0.35 $\pm$ 0.29 \\
      ZZZ & Ac &  NH$_2$ &$\omega_3$ & 0.16 $\pm$ 0.23 \\
    \end{tabular}
    
    \label{tab:my_label}
\end{ruledtabular}
   
\end{table}


\subsection{Conformational Landscapes and Supramolecular Aggregation}\label{subsec:assembly}

Our third test sought to assess the capacity of MoSiC-CGenFF-NTOID to predict experimentally-observed trends in peptoid self-assembly. This assessment goes beyond the single-molecule validations conducted thus far, and where successful prediction of assembly preferences is a prerequisite for the design of functional peptoid nanomaterials and exposure of underlying sequence-dependent design rules.
The preponderance of experimental peptoid self-assembly studies consider long-chain peptoids, the assembly behavior of which is challenging to simulate due to the long characteristic time scales associated with forming ordered supramolecular nanostructures \cite{Zhao2020, Zhao2022, Zhao2023, Mannige2015, Robertson2016}. One exception is the recent study by Castelletto et al.\ that considered the assembly of minimal peptoid sequences based on analogs of peptide Phe and Lys residues \cite{Castelletto2020}. Specifically, this work considered the assembly in water of four heterotrimers -- NPhe--Nae--NPhe (FXF), NPhe--NLys--NPhe (FKF), Nae--NPhe--NPhe (XFF), and NLys--NPhe--NPhe (KFF) -- and the resulting assemblies analyzed by cryo-TEM. This analysis revealed only the FXF sequence to form linear nanowires whereas the remaining three formed globular structures (Figure \ref{fig:pca-expt}A). Similar to peptide nanowires, peptoid nanowires could be employed in applications such as biomimetic electronics\cite{Creasey2019}, nanoparticle joining\cite{Kaur2010-zk}, and substrates for cell culture\cite{Walls2020}. Moreover, the distinct structural features and functional groups of peptoids could potentially expand their versatility and enable novel functional applications. We hypothesized that the MoSiC-CGenFF-NTOID model should be able to recapitulate the results of these experiments by distinguishing the different self-assembly behaviors of these four peptoid sequences. To test this hypothesis, we examined the structures of these peptoids on two levels: first, the statistics of a single chain, and second, the aggregation behavior of a large number of chains. 

\subsubsection{Single-Chain Secondary Structure} \label{subsec:assembly_single}

\begin{figure*}
     \centering
     \includegraphics[width=\textwidth]{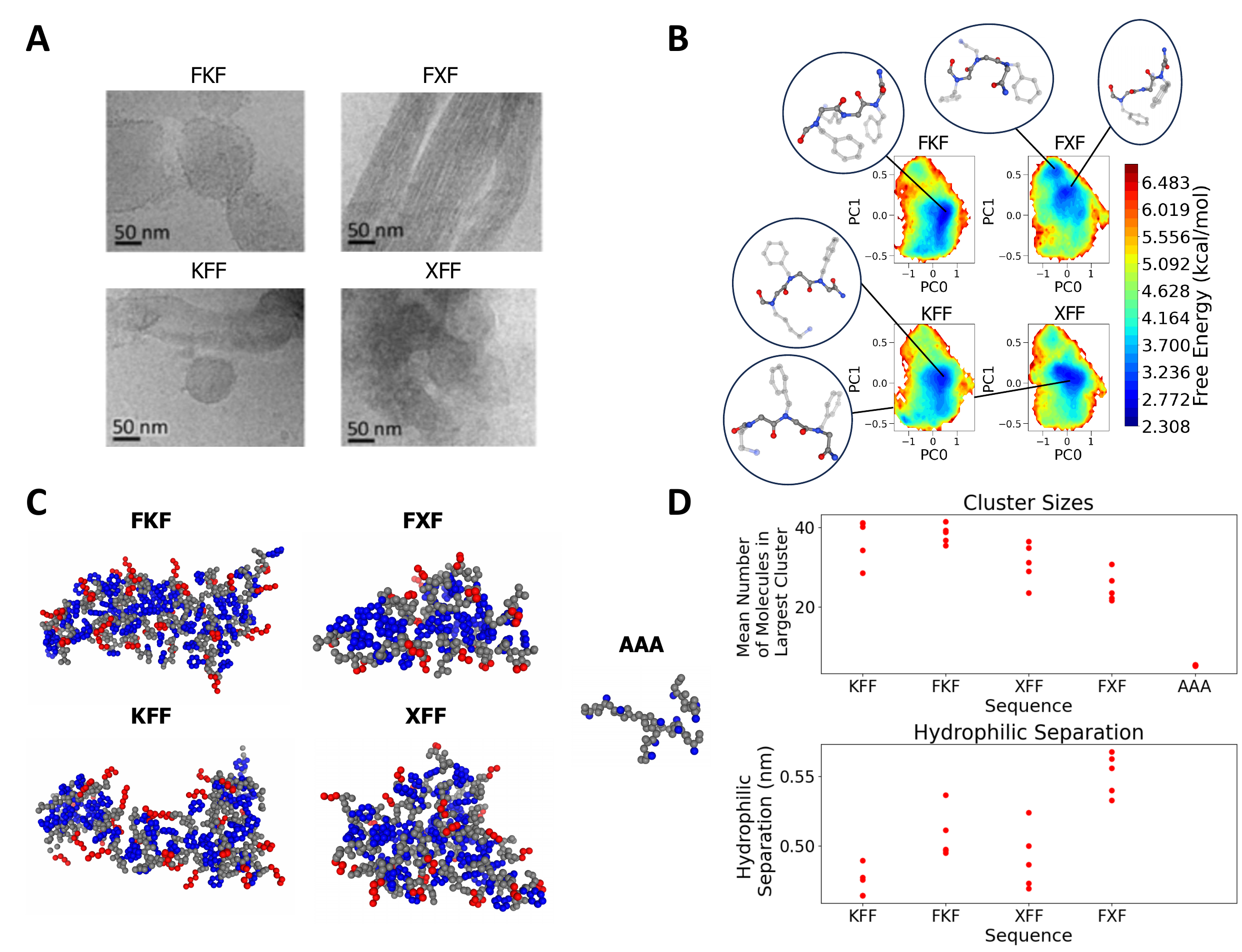}
     \caption[skip=0pt]{Assessment of the capabilities of MoSiC-CGenFF-NTOID to predict the single-chain statistics and self-assembly behavior of minimal peptoid trimers FKF, FXF, KFF, and XFF. (A) Cryo-TEM images of the self-assembled structures formed in water and reported by Castelletto et al.\ \cite{Castelletto2020}. The FXF sequence forms ordered nanofibers whereas the remaining sequences produce less ordered globular aggregates. Reprinted from Ref.~\cite{Castelletto2020} under a Creative Commons Attribution (CC-BY) license. (B) Free energy surfaces for the four peptoids projected into the leading two principal components [PC0, PC1] from principal components analysis of the ensemble of trajectories of the 34 peptoid homotrimers. Representative structures corresponding to the observed free energy minima are displayed around the periphery. The FXF sequence exhibits a free energy minimum in the upper-left of the landscape unpopulated by the other sequences and containing collapsed molecular structures with a curled backbones. (C) Space-filling visualization of representative large aggregates observed in simulations of 50 peptoid chains in water constructed using NGLView\cite{10.1093/bioinformatics/btx789}. Hydrophobic side chains are colored blue, positively charged side chains are colored red, and the backbone colored gray. The FXF sequence tends form smaller clusters than FKF, XFF, and KFF, and the NPhe side chains tend to be oriented towards the core of the cluster with the backbone and Nae side chains located on the periphery. The trisarcosine (AAA) control shows extremely limited aggregation propensity. (D) Quantification of the mean number of molecules in the largest extant cluster in each frame recorded in the 50 chain self-assembly simulation trajectories (upper plot) and the ``hydrophilic separation'' -- the nearest neighbor inter-molecular distance between any two non-NPhe backbone nitrogens for chains within a cluster (lower plot). The hydrophilic separation for AAA is undefined since it possesses no hydrophilic residues. Of the four sequences, FKF, FXF, KFF, and XFF, the FXF sequence exhibits smaller clusters with more distantly separated chain backbones. 
     }
     \label{fig:pca-expt}
\end{figure*}

We compare the single-chain statistics by constructing and comparing the free energy landscapes for the four chains:  Ac--NPhe--Nae--NPhe--NHMe (FXF), Ac--NPhe--NLys--NPhe--NHMe (FKF), Ac--Nae--NPhe--NPhe--NHMe (XFF), and Ac--NLys--NPhe--NPhe--NHMe (KFF). The experiments were conducted at pH 3, under which conditions the terminal amino groups on the Nlys and Nae side chains are expected to exist in the \ce{-NH3+} protonated state, and we match these ionization patterns in our calculations. Contrary to the experiments, we simulate acetyl N-capped and aminomethyl C-capped sequences that we model as electrically neutral. We define a data-driven low-dimensional basis appropriate for exposing the conformational landscapes of peptoid trimers by applying principal components analysis (PCA) to our WT-PBMetaD simulations of the 34 acetyl N-capped and aminomethyl C-capped homotrimers. Specifically, we constructed a rototranslationally invariant and sequence agnostic featurization of each trajectory using the pairwise distances between the heavy backbone atoms and extracting the leading two principal components, PC0 and PC1, that together explain 62.8\% of the variance in the data. We present in Figure \ref{fig:pca-expt}B the free energy landscapes for the FXF, FKF, XFF, and KFF heterotrimers computed by reweighting and projecting the WT-PBMetaD trajectories into PC0 and PC1. For comparison, we present analogous free energy landscapes of all 34 peptoid homotrimers in \blauw{Figure S4}. We also generate physical insight into the two leading PCs, we correlated them with candidate physical variables and observed strong correlation of PC0 with the molecular radius of gyration $R_g$ (\blauw{Figure S5}) and PC1 with the cis/trans state of the dihedral angle between the first and second residues, represented by $\cos \omega_2$ (\blauw{Figure S6}).

The free energy landscapes in Figure \ref{fig:pca-expt}B expose relatively similar conformational ensembles for the three sequences FKF, KFF, and XFF, dominated by a primary free energy minimum at [PC0$\approx$0.2, PC1$\approx$0.1] comprising molecular conformations with extended backbones. The free energy landscape for FXF differs in that the dominant minimum is shifted to [PC0$\approx$-0.1, PC1$\approx$0.3], corresponding to an elevated cis preference in $\omega_2$, and the presence of a secondary minimum at [PC0$\approx$-0.4, PC1$\approx$0.5] containing conformations with collapsed (i.e., low-$R_g$) molecular configurations with a curled backbone. Given that FXF is the only one of the four sequences experimentally observed to form ordered nanofibers (Figure \ref{fig:pca-expt}A), we hypothesize that these conformations uniquely populated by FXF at the level of a single chain play a role in mediating the self-assembly of the FXF trimer into nanowires.

\subsubsection{Multi-Chain Aggregation} \label{subsec:assembly_multi}

Next, we tested the capacity of our model to predict the self-assembly behavior of the four sequences in multi-chain simulations. We performed simulations of 50 chains of FKF, FXF, KFF, and XFF heterotrimers in water along with a trisarcosine AAA control that is known not to show strong aggregation behaviors due to the high solubility of polysarcosine relative to other polypeptoids\cite{BIRKE2018163}. Due to the intractably high computational expense associated with biasing all dihedral angles within these multi-chain simulations, enhanced sampling was not employed, but rather initial conformations of each chain were rather generated with $\omega$ dihedral isomeric states sampled from the cis/trans ratios determined in our single-chain simulations and the system evolved under unbiased molecular dynamics. A deficiency of this approach is that the high free energy barriers associated with cis/trans transitions mean that the conformational state of the molecules are likely locked into their initial conditions, but are constructed to obey the single chain statistics. 

The FKF, FXF, KFF, and XFF simulations exhibited self-assembly behaviors that produced long-lived clusters whereas the AAA sequence evinced only transient intermolecular associations. The cluster distribution attained at 20 ns persisted approximately unchanged for the remainder of the 300 ns simulation. Cluster analysis was performed using the NetworkX Python package\cite{SciPyProceedings_11}. Each molecule was treated as a node and an edge was constructed for intermolecular heavy atom distances less than 3.5 \r{A}. A cluster was defined as a connected clique of three or more nodes. Appreciating the non-equilibrium and history-dependent nature of cluster formation on these time scales, analyses were conducted over for five independent 300 ns trajectories, in each case discarding the initial 20 ns transient. 

In Figure \ref{fig:pca-expt}C we present molecular visualizations of representative clusters for each system. Visual analysis exposes a number of apparent differences between the five sequences. The AAA control produces only small and transiently associated clusters. The FKF, FXF, KFF, and XFF sequences all produce clusters ranging in size from 3-50 molecules, with individual molecules dynamically associating and dissociating within these clusters over the course of the simulation. The large clusters expose subtle, but clearly visible, differences in the self-assembled cluster architectures. The FKF, KFF, and XFF sequences tend to produce clusters with the NPhe side chains and peptoid backbones constituting a hydrophobic core surrounded by a corona of positively-charged NLys or Nae side chains exposed to the water solvent. There also appears to be no strong directional preference for the NPhe side chains. The FXF sequence tends to form smaller clusters with a slightly altered architecture wherein the NPhe side chains are preferentially oriented towards the center of the cluster with the peptoid backbones and Nae side chains removed to the periphery.


We supplemented these qualitative observations with quantitative analyses of the cluster size and organization. We calculated the mean number of molecules in the largest cluster in each frame over the course of the five self-assembly simulation trajectories (Figure \ref{fig:pca-expt}D, upper plot). Consistent with our visual analysis, the mean number of molecules in the largest cluster was significantly smaller for FXF (24.9 $\pm$ 1.7) compared to FKF (38.4 $\pm$ 1.0), KFF (37.0 $\pm$ 2.5), and XFF (31.0 $\pm$ 2.3), where uncertainties represent standard errors computed over the five independent runs. We then computed the nearest neighbor inter-molecular distance between any two non-NPhe backbone nitrogens for chains within a cluster as a measure of the proximity of two chains. We define this distance based on the NLys or Nae backbone nitrogens to emphasize the proximity of the hydrophilic components of the chains and term this distance the ``hydrophilic separation''. This definition was motivated by the observation of elevated partitioning and orientation of the FXF NPhe side chains towards the core of the cluster, and it was our expectation that the hydrophilic separation should provide a good metric by which to characterize this organizational difference between the sequences. The mean hydrophilic separation for FXF is (0.552 $\pm$ 0.007) nm compared to  (0.507 $\pm$ 0.008) nm, (0.474 $\pm$ 0.005) nm, and (0.491 $\pm$ 0.010) nm for FKF, KFF, and XFF, respectively (Figure \ref{fig:pca-expt}D, lower plot). This is consistent with the visual organization of the clusters and quantifies the closer packing and aromatic associations of the NPhe side chains in the FXF cluster cores. The different patterns in cluster size and architecture observed for FXF in our small simulations are consistent with the experimental observations of a different self-assembled architecture (i.e., nanowires) for FXF compared to FKF, KFF, and XFF (i.e., globules) (Figure \ref{fig:pca-expt}A) \cite{Castelletto2020}. The elevated NPhe associations observed within FXF are also consistent with the characterization by Castelletto et al.\ that $\pi$-stacking is likely a predominant molecular driving force for the formation of the nanowire geometries 
The capacity of the MoSiC-CGenFF-NTOID model to identify FXF as an outlier relative to FKF, KFF, and XFF in both single chain statistics and self-assembly behavior in line with experimental observation of self-assembly trends provides support that the model is capable of predicting aggregation trends in multi-chain peptoid systems.

\section{Computational Design of Self-Assembling Peptoid Sequences}

After demonstrating a retrospective consistency of the MoSiC-CGenFF-NTOID predictions with the self-assembly patterns within the family of FKF, FXF, KFF, and XFF minimal peptoid sequences, we now sought to extrapolate the principles exposed within this analysis to the prospective discovery of additional peptoid sequences with similar self-assembly character to FXF. It was our hypothesis that sequences exhibiting similar single-chain statistics and multi-chain assembly character may also be capable of forming nanowires in experimental tests. Commencing from the FXF heterotrimer, we conducted a virtual screen of 11 mutants constructed by either replacing the central Nae residue for an alternative charged or polar residue (FRF, FEF, FDF, FOF, FCF, FNF, FSF, and FTF) or by replacing the two flanking NPhe residue for a bulky hydrophobe (WXW, LXL, and ZXZ). This ensemble represents a rationally-designed family of heterotrimers with similar physicochemical properties to FXF that we conjecture may also exhibit similar assembly behaviors. Of course with sufficient computational resources, the virtual screen may be scaled up to consider all 34\textsuperscript{3} = 39,304 possible trimers containing all possible combinations of the 34 side chains currently supported by the model.

\begin{figure}
    \centering
    \includegraphics[scale=0.36]{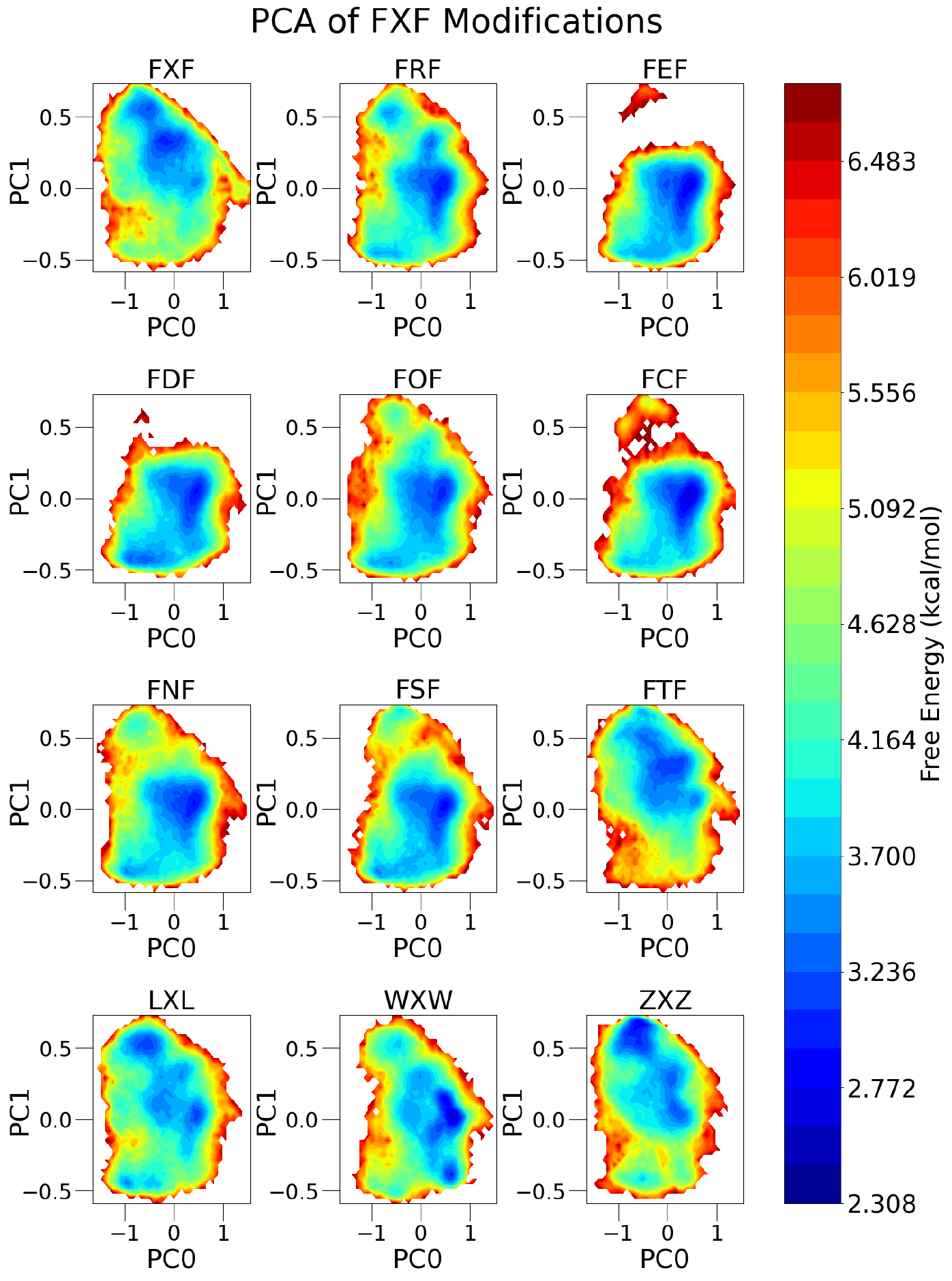}
    \caption{Free energy surfaces of FXF and 11 mutants peptoids projected into the leading two principal components [PC0, PC1] from principal components analysis of the ensemble of trajectories of the 34 peptoid homotrimers. Mutations either exchange the central Nae residue for a polar, or charged moiety (FRF, FEF, FDF, FOF, FCF, FNF, FSF, and FTF), or the two flanking NPhe residues for a bulky hydrophobe (LXL, WXW, and ZXZ).}
    \label{fig:prediction}
\end{figure}

Considering first the single-chain statistics, we present in Figure \ref{fig:prediction} the [PC0,PC1] free energy surfaces calculated for the 11 mutants and FXF parent using the approach detailed in Section \ref{subsec:assembly_single}. Of these, we identified six mutant sequences -- FTF, FRF, FOF, LXL, WXW, and ZXZ -- that display substantially populated free energy minima in the upper-left regions of the plot exhibiting significant overlap with those of FXF. We hypothesized that adoption of a similar single-chain conformational ensemble to FXF may promote similar aggregation behaviors in these sequences.


\begin{figure}
    \centering
    \includegraphics[scale=0.33]{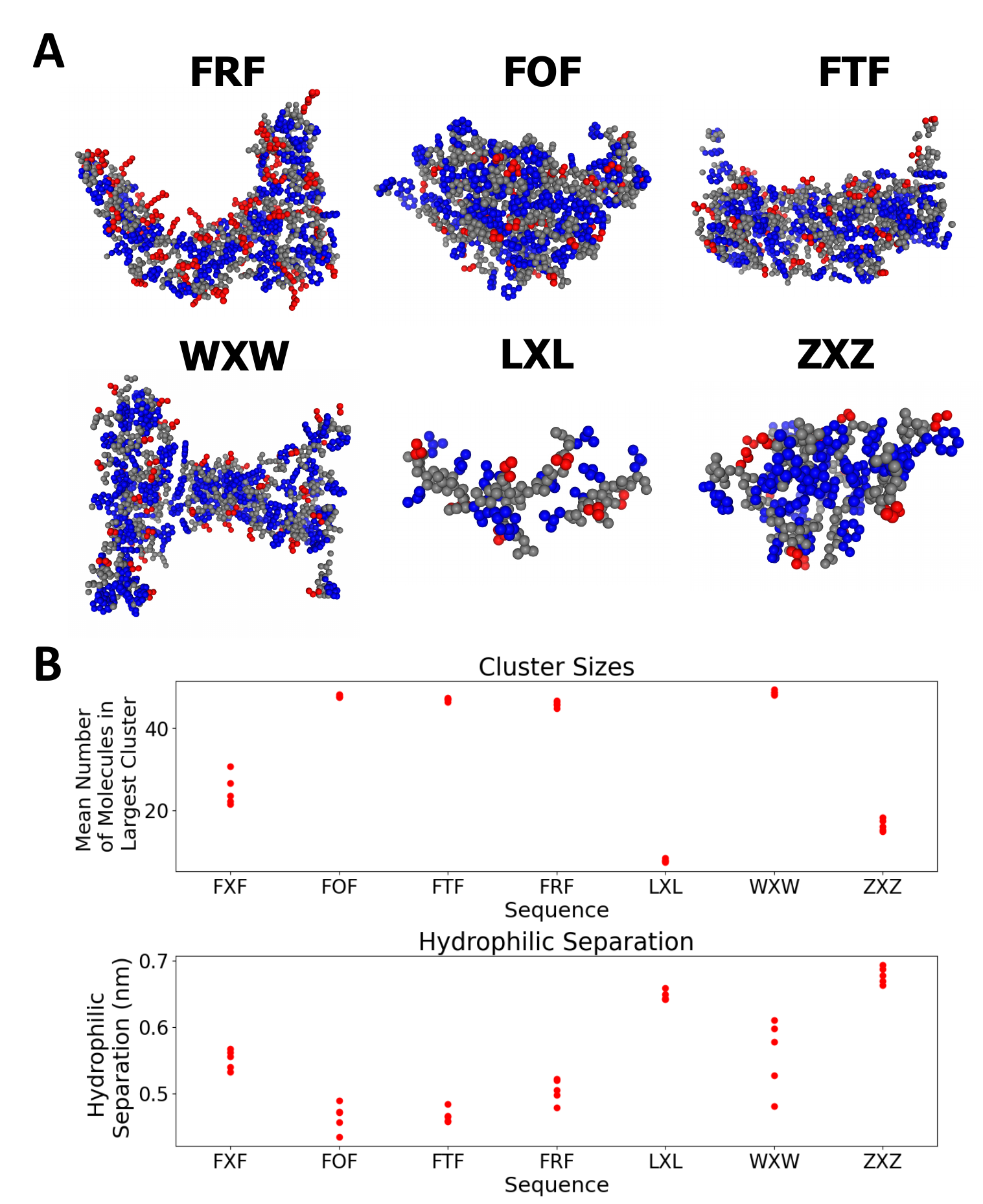}
    \caption{Computational screening and design of peptoid trimers to promote nanowire assembly behaviors. (A) Space-filling visualization of representative large aggregates observed in simulations of 50 peptoid chains in water constructed using NGLView\cite{10.1093/bioinformatics/btx789}. Hydrophobic side chains are colored blue, hydrophilic side chains are colored red, and backbones are colored gray. (B) Quantification of the mean number of molecules in the largest cluster (upper plot) and the hydrophilic separation (lower plot).}
    \label{fig:newclusters}
\end{figure}

We next subjected these six sequences to multi-molecular simulations of 50 peptoid chains in water using the same protocol detailed in Section \ref{subsec:assembly_multi}. Representative large clusters from these calculations are presented in Figure \ref{fig:newclusters}A, while an accounting of the observed cluster sizes and hydrophilic separations can be found in Figure \ref{fig:newclusters}B. These results show that when the central Nae (X) residue is substituted for Noe (O), NThr (T), or NArg (R), the resulting sequences -- FOF, FTF, and FRF -- tend to form aggregates with similar cluster sizes ($>$35), hydrophilic separations ($<$0.52 nm), and visual character to the KFF, FKF, and XFF sequences that were experimentally observed to form globular, as opposed to nanowire, aggregates (Figure \ref{fig:pca-expt}). Conversely, when the exterior NPhe (F) residues are substituted for NLeu (L), NTrp (W), or Nspe (Z), the resulting sequences -- LXL, WXW, and ZXZ -- tend to form clusters with large hydrophilic distances equal or greater than that of FXF ($\geq$0.552 nm). Maximum cluster sizes for ZXZ ($16.4 \pm 0.6$) are more comparable to those of FXF ($24.9 \pm 1.7$), whereas WXW forms much larger clusters incorporating nearly all 50 chains in the box, and LXL forms transient clusters of very small size ($7.7 \pm 0.2$). The morphology of the WXW and ZXZ clusters is more visually similar to FXF, with the NPhe side chains localized and oriented towards the center of the aggregate, whereas the LXL clusters are rather more diffuse with the NLeu side chains exposed to solvent.

Taken together, this small-scale virtual screen of mutations localized around FXF suggests that single-chain statistics are insufficient to predict the multi-chain aggregation behavior of these minimal peptoids, but that it can identify a number of sequences for more expensive multi-chain simulations that do possess similar behaviors. In particular, the ZXZ and, to some degree WXW, sequences possess similar aggregation patterns and morphologies to FXF, and we suggest that they may warrant further exploration as promising candidates for nanowire self-assembly. Moreover, our screen indicates the seemingly dual requirement for a central Nae (X) residue coupled with bulky aromatic exterior residues (F, Z, W) to promote the desired assembly patterns.

\section{Conclusions}

In this work, we present MoSiC-CGenFF-NTOID as a new force field for all-atom simulation of peptoids that makes provision for diverse side chain chemistries in a modular and extensible fashion. The heart of the approach is the hypothesis that side chains within the large family of substituted methyl groups (i.e., \ce{-CH_3}, \ce{-CH_2R}, \ce{-CHRR$^{\prime}$} \ce{-CRR$^{\prime}$R$^{\prime\prime}$}) can be accurately modeled using the CGenFF-NTOID parameterization of Weiser and Santiso for the backbone (i.e., CGenFF with refitted $\rho$, $\psi$, and $\omega$ backbone dihedrals) \cite{https://doi.org/10.1002/jcc.25850} and default CGenFF parameters for the side chain \cite{Vanommeslaeghe2010-ao}. Under this modular hypothesis, any substituted methyl side chain parameterizable by CGenFF can be extensibly incorporated into the force field. We refer to this model and paradigm as Modular Side Chain CGenFF-NTOID (MoSiC-CGenFF-NTOID). We validate the force field in three primary evaluations: a comparison of Ramachandran potential energy landscapes to \textit{ab initio} calculations, cis/trans isomerization energies and free energies against \textit{ab initio} calculations and experimental measurements, and predictions of experimentally observed peptoid self-assembly. We make the MoSiC-CGenFF-NTOID force field, associated peptoid structure generator, and instructions on how to incorporate additional side chains beyond those currently supported freely available to the community as an open-source package available at \url{https://github.com/UWPRG/mftoid-rev-residues}.

While we advise validating the parameterization of any new side chain by, at a minimum, comparing Ramachandran potential energy plots against those generated by \textit{ab initio} calculations to determine whether additional $\chi_1$ dihedral reparameterization may be warranted using the procedure developed by Weiser and Santiso\cite{https://doi.org/10.1002/jcc.25850}, our calculations have provided strong support for the modular hypothesis wherein all 31 side chains tested that are members of the family of substituted methyl groups were ranked as performing good or fair under our evaluation metrics. Two of the three side chains that are not substituted methyl groups -- Gly and Pro -- are identical to their peptide analogs and also performed well due to the existence of prior parameterizations within CGenFF. The third non-substituted methyl side chain, Nph, was constructed as a negative control and the only one to exhibit poor performance. This extensibility is an extremely valuable attribute of a peptoid force field due to the essentially infinite chemical variability in possible side chains. In this regard, we hope that MoSiC-CGenFF-NTOID may prove valuable to the community in opening the door to high-throughput virtual screening efforts for the computational evaluation, design, and engineering of novel peptoid materials, without requiring laborious reparameterization of each new side chain chemistry. In particular, we see opportunities for couplings to machine learning, active learning, and Bayesian optimization paradigms to engineer desired secondary, tertiary, and quaternary structures of peptoid chains, assemblies, and functional materials, including peptidomimetic enzymes and novel therapeutics.

\section{Data Availability}

The MoSiC-CGenFF-NTOID force field, associated peptoid structure generator, and instructions on how to incorporate additional side chains beyond the 34 currently supported is freely available from \url{https://github.com/UWPRG/mftoid-rev-residues}.

\section{acknowledgements}

This material is based upon work supported by the US Department of Energy (DOE), Office of Science, Office of Basic Energy Sciences (BES), as part of the Energy Frontier Research Centers program: CSSAS -- The Center for the Science of Synthesis Across Scales -- under Award Number DE-SC0019288. This work was completed in part with resources provided by the University of Chicago Research Computing Center. We gratefully acknowledge computing time on the University of Chicago high-performance GPU-based cyberinfrastructure supported by the National Science Foundation under Grant No.\ DMR-1828629.

\section*{Conflict of Interest Disclosure}

A.L.F.\ is a co-founder and consultant of Evozyne, Inc.\ and a co-author of US Patent Applications 16/887,710 and 17/642,582, US Provisional Patent Applications 62/853,919, 62/900,420, 63/314,898, 63/479,378, 63/521,617, and 63/669,836, and International Patent Applications PCT/US2020/035206, PCT/US2020/050466, and PCT/US24/10805.




\clearpage
\newpage

\bibliography{bibliography}
\bibliographystyle{unsrt}

\end{document}